\documentclass[journal]{IEEEtran}
\usepackage{bbm}
\usepackage{cite}
\usepackage{amsfonts}
\usepackage{mathrsfs}
\usepackage{graphicx}
\usepackage{amssymb}
\usepackage{latexsym}
\usepackage{amsmath}
\usepackage{stfloats}
\usepackage{cases}
\usepackage{setspace}
\usepackage{bigstrut}
\usepackage{bm}
\usepackage{url}
\usepackage{color}
\usepackage{subfigure}
\usepackage[ruled,linesnumbered]{algorithm2e}
\usepackage{amsthm}

\begin{document}

\title{Distributed Near-Field Channel Estimation for

U6G XL-MIMO Systems under Beam Squint}
    \author{Zhizheng Lu,~\IEEEmembership{Student Member, IEEE}, Yu Han,~\IEEEmembership{Member, IEEE}, Xiao Li,~\IEEEmembership{Member, IEEE},
    
    Shi Jin,~\IEEEmembership{Fellow, IEEE}, and Michail Matthaiou,~\IEEEmembership{Fellow, IEEE}
    
    \thanks{Manuscript received April 03, 2025; revised August 21, 2025; accepted November 10, 2025. This work was supported in part by the National Natural Science Foundation of China (NSFC) under Grants 62422105 and 62261160576, in part by the Key Technologies RD Program of Jiangsu (Prospective and Key Technologies for Industry) under Grants BE2023022, BE2023022-1 and BE2023022-2, and in part by the Natural Science Foundation of Jiangsu Province under Grant BK20230824. The work of M. Matthaiou was supported by the European Research Council (ERC) under the European Union’s Horizon 2020 research and innovation programme (grant agreement No. 101001331). \textit{(Corresponding author: Shi Jin, Yu Han.)}}
    
    \thanks{Z. Lu, Y. Han, X. Li, and S. Jin are with the National Mobile Communications Research Laboratory, Southeast University, Nanjing 210096, China (e-mail: luzz@seu.edu.cn; hanyu@seu.edu.cn; li\_xiao@seu.edu.cn; jinshi@seu.edu.cn).}
    
    \thanks{M. Matthaiou is with the Centre for Wireless Innovation (CWI), Queen’s University Belfast, BT3 9DT, Belfast, U.K. (e-mail: m.matthaiou@qub.ac.uk).}
    
    \thanks{This is an extended and revised version of a previous conference paper that appeared at the \textit{IEEE WCNC 2025} \cite{conference}.}}
\maketitle

\begin{abstract}

Since the beam squint and near-field effects both inherently exist in upper-6 GHz (U6G) extremely large-scale multiple-input multiple-output (XL-MIMO) systems, wideband near-field channel estimation faces severe challenges, such as higher computational complexity, and higher pilot overhead particularly at hybrid architectures with fewer radio frequency (RF) chains. To precisely reduce the complexity and number of pilots, the \textit{parametric symmetry of wideband near-field channels} is explored, such that the channel parameters, including angle, distance, and range, can be decoupled based on the delay variations observed by different antennas. Based on this, a \textit{distributed parametric symmetry-based (DPS) algorithm}, applicable to U6G XL-MIMO, is proposed. The delays observed by different subarrays are estimated and extrapolated across the local processing units (LPUs) firstly, and then, the channel parameters are decoupled and estimated at the central processing unit (CPU), by only linearly combining the delays from different LPUs. The path gains are calculated at different LPUs, respectively, to reconstruct the channel with low complexity. Since the proposed algorithm does not rely on scanning the polar-domain dictionary, only \textit{a single pilot} is required even with hybrid architectures. Furthermore, the computational complexity, multiple-path resolution, Cramér–Rao lower bound (CRLB) and lower bound (LB) of the estimates in hybrid architectures and the DPS algorithm, respectively, are analyzed, to evaluate the realizable potential of the proposed algorithm. The simulation results prove that the proposed algorithm has a higher estimation accuracy, while requiring less complexity and pilots.

\end{abstract}

\section{Introduction}\label{Sec: Introduction}

Extremely large-scale multiple-input multiple-output (XL-MIMO) is perceived as a promising technology for the sixth generation (6G) wireless systems \cite{XL-MIMO}, to satisfy the rapidly increasing demands on spectral efficiency (SE) and coverage \cite{6G1}. Through employing a much larger number of antennas at future base stations (BSs), an improved quality of service for user equipment (UEs) can be provided \cite{6Gperformance}. Moreover, the uppper mid-band spectrum, such as the upper-6 GHz (U6G) frequency band (6.425 GHz-7.125 GHz), is emerging as a strategic choice for international mobile telecommunications (IMT) services in 6G \cite{U6G}, since it can provide abundant spectrum resources, and can also avoid the challenge of high path loss in the millimeter wave (mmWave) and terahertz (THz) frequency bands. Meanwhile, the higher frequency compared to the sub-6 GHz means that the size of antennas can be further reduced, and thus a larger number of antennas can be deployed within a limited area. Therefore, U6G XL-MIMO systems showcase strong potential in future wireless communication.

However, the large number of antennas will inevitably increase the Rayleigh distance \cite{Rayleighdistance}, and will lead to the near-field effect for XL-MIMO systems \cite{near-field3}. In this case, the phase variations caused by different transmission distances across the entire array must be considered, while more channel parameters need to be estimated in the near-field region than in the far-field region, which will entail a larger computational complexity for channel state information (CSI) acquisition. Moreover, the wide bandwidth in the U6G frequency band will lead to a non-negligible beam squint effect for XL-MIMO systems \cite{beamsquinteffect1,beamsquinteffect3,beamsquinteffect4}.\footnote{The beam squint effect is also known as the spatial and frequency dual-wideband effect \cite{defineofdualwideband}, and is caused by the wide bandwidth and large array simultaneously.} Consequently, the phase variations caused by different frequencies across all the subcarriers also need to be considered, which will undermine the spatial-domain sparsity of near-field channels, and will bring critical challenges for wideband near-field channel estimation in U6G XL-MIMO systems, such as higher computational complexity and low CSI estimation accuracy.

\subsection{Prior Works}

There have been many studies on channel estimation considering the near-field and the beam squint effects. In particular, \cite{near-field_CE1,near-field_CE2,near-field_CE3,near-field_CE4,near-field_CE5} studied the near-field channel estimation for XL-MIMO systems. The work of \cite{near-field_CE1} modeled the realistic phase variations of the received signal, and studied the polar-domain sparsity of the near-field channel. Then, an off-grid polar-domain simultaneous iterative gridless weighted algorithm was proposed to detect the channel parameters and reconstruct the near-field channel. The authors of \cite{near-field_CE2} conceived a more realistic near-field channel model, which considered the realistic amplitude and phase variations simultaneously, and further improved the estimation accuracy when the distance is short. Moreover, \cite{near-field_CE3} considered a reconfigurable intelligent surface (RIS)-aided mmWave system in the near-field region, and proposed a single-path cascaded channel estimation algorithm based on the Toeplitz covariance matrix. To decrease the computational complexity, \cite{near-field_CE4} proposed a novel extremely large-scale RIS (XL-RIS) architecture, which can decouple the estimation procedure of the angle and distance parameters. Then, an efficient three-stage scheme was also proposed to reconstruct the cascaded near-field channel. The work of \cite{near-field_CE5} focused on the hybrid-field scenario, and proposed a hybrid-field channel estimation algorithm, in which the near-field paths were detected and removed firstly, and then the far-field paths were estimated from the residual signal.

The works in \cite{beamsquint1,beamsquint2,TTDsystem0,TTDsystem1,TTDsystem2,utilizebeamsquint,wideband_CE1,wideband_CE2,wideband_CE3,wideband_CE4,wideband_CE5,wideband_CE6} considered the influence of the beam squint in mmWave and THz wideband wireless systems. The performance degradation caused by the beam squint was studied in \cite{beamsquint1,beamsquint2}, which showed that the SE will decrease when the bandwidth is large. Then, the authors in \cite{TTDsystem0,TTDsystem1,TTDsystem2} proposed a beamforming architecture based on true-time-delay lines (TTDs) to eliminate the beam squint effect. The work in \cite{utilizebeamsquint} found that the beam squint effect can be freely controlled through the TTDs, which will make it possible to reversely utilize the beam squint effect for UE localization with low overhead. Moreover, the works in \cite{wideband_CE1,wideband_CE2,wideband_CE3} studied the channel estimation problem for wideband systems in the far-field region. The authors in \cite{wideband_CE1} explored the shift-invariant block sparsity of the far-field channel, and also proposed an effective block sparsity-based algorithm for channel estimation. The authors in \cite{wideband_CE2} used a densely-spaced antenna structure and consecutive sub-carrier assignment approach, which can effectively avoid the aliasing effect and reduce the ambiguity during the initialization stage to decrease the pilot overhead. The authors in \cite{wideband_CE3} considered the correlation of estimates between adjacent subcarriers, and then jointly detected the sparse support with frequency-dependent sparse structure to improve the estimation accuracy. Furthermore, based on the modified wideband discrete linear chirp transform, the work in \cite{wideband_CE4} proposed a low-complexity sparse multiple-path near-field channel estimation algorithm. Near-field channel estimation for XL-RIS-aided wideband systems was studied in \cite{wideband_CE5}, in which a wideband polar-domain dictionary was constructed, and a correlation coefficient-based atom matching method was presented to recover the cascaded channel. The authors in \cite{wideband_CE6} studied the hybrid-field wideband channel estimation under beam squint, in which linearly corrected angle-domain and polar-domain dictionaries were designed to estimate the far-field and near-field channels, respectively.

However, due to the high-dimensional received signals and the limited number of RF chains in wideband XL-MIMO systems, the existing works on channel estimation rely on polar-domain dictionaries and, as such, still require high complexity and an excessive number of pilots. Hence, how to reduce the complexity and pilots by exploring wideband near-field features brought by the beam squint effect, is a very timely question for U6G XL-MIMO systems. Moreover, new channel features in the U6G frequency bands, such as more paths than in the mmWave frequency bands, while the spatial sparsity still exists, should also be considered. To the authors’ best knowledge, the relevant research is still lacking.

\subsection{Contributions}

In this study, a U6G wideband XL-MIMO system with a distributed hybrid architecture is considered. The \textit{parametric symmetry} brought by the beam squint effect in the near-field channel is explored firstly. Then, a \textit{distributed parametric symmetry-based (DPS) algorithm} entailing only \textit{a single pilot} is proposed, to estimate the wideband near-field channel effectively. The delays of different subarrays are estimated and extrapolated across local processing units (LPUs), and then the channel parameters can be decoupled and estimated at the central processing unit (CPU). Finally, the path gains are calculated at different LPUs to reconstruct the full-dimensional channel. Compared with other works, the major contributions of this paper are summarized as follows:

\begin{itemize}
\item The parametric symmetry of the wideband near-field channel, which is caused by the beam squint and near-field effects, is leveraged. The wideband near-field channel still has the delay-domain sparsity if we observe it from each antenna individually. When the delay of each antenna is obtained, the near-field channel parameters can be decoupled through a linear combination, underpinning the potential ability to decrease the computational complexity in channel estimation.
\end{itemize}

\begin{itemize}
\item By availing of the parametric symmetry, a DPS algorithm is proposed to estimate the wideband near-field channel effectively, which involves only a single pilot. The delays of different subarrays are estimated and extrapolated across the LPUs, and then the CPU combines the delays from the LPUs, to decouple and estimate the channel parameters with much low computational complexity. Based on this, the complex gains are calculated by the LPUs to reconstruct the channel.
\end{itemize}

\begin{itemize}
\item The computational complexity, multiple-path resolution, Cramér–Rao lower bound (CRLB) and lower bound (LB) of the estimates in hybrid architectures and the proposed algorithm, respectively, are analyzed, to evaluate the estimation performance under the near-field and beam squint effects. The proposed DPS algorithm has LBs of estimates close to the CRLBs, despite the lower number of pilots and lower complexity. The simulation results verify that the proposed DPS algorithm exploiting the parametric symmetry has a higher estimation accuracy than that of other methods, while it entails only a single pilot with reduced computational complexity.
\end{itemize}

\subsection{Organization and Notation}

The rest of this paper is organized as follows: In Section $\rm \uppercase\expandafter{\romannumeral2}$, the hybrid XL-MIMO systems with distributed processing architectures and the U6G wideband near-field channel model are introduced. In Section $\rm\uppercase\expandafter{\romannumeral3}$, the parametric symmetry in wideband near-field channels is explored, and then the DPS algorithm is proposed to estimate the channel. In Section $\rm \uppercase\expandafter{\romannumeral4}$, the numerical analysis of the proposed algorithm is provided. In Section $\rm \uppercase\expandafter{\romannumeral5}$, the simulation results are presented to evaluate the proposed DPS algorithm. Finally, our conclusions are given in Section $\rm \uppercase\expandafter{\romannumeral6}$.

\textit{Notation:} Henceforth, lower-case and upper-case bold letters denote vectors and matrices, respectively; $\odot$ and $\otimes$ represent the Hadamard product and Kronecker product. For a matrix ${\bf{A}}$, ${{\bf{A}}^{\rm{T}}}$, ${{\bf{A}}^{\rm{H}}}$ and ${{\bf{A}}^{\dagger}}$ represent the transpose, conjugate transpose and pseudo-inverse operations, respectively; $\left[\bf A\right]_{m,:}$, $\left[\bf A\right]_{:,n}$, and $\left[\bf A\right]_{m,n}$ are the $m$-th row, $n$-th column, and $\left(m,n\right)$-th element of $\bf A$; $\left[{\bf a}\right]_n$ represents the $n$-th element of vector ${\bf a}$; ${\rm blkdiag} \left({\bf a}_1,\ldots,{\bf a}_{N}\right)$ represents a block diagonal matrix that deploys the vector blocks $\left\{ {\bf a}_1,\ldots,{\bf a}_N\right\}$ in the diagonal, while ${\bf 1}_{N}$ represents the $N$-dimensional all-one vector. Moreover, $\left\lfloor {\cdot} \right\rfloor$ and $\left\lceil {\cdot} \right\rceil$ are the round-down and round-up operations; $\left| \cdot \right|$ and $\left\| \cdot \right\|$ represent taking the absolute value operation and the vector ${l_2}$ norm, respectively; $\mathbb{E} \left\{ \cdot \right\}$ denotes the statistical expectation; ${\cal R} \left\{ {\bf A} \right\}$ represents the real part of ${\bf A}$; ${\rm vec}\left\{ \cdot\right\}$ is the matrix vectorization operation, and ${\cal C}{\cal N} \left( {{\boldsymbol\mu} ,{\bf{\Sigma }}} \right)$ denotes the complex Gaussian distribution with mean $\boldsymbol\mu $ and covariance ${\bf{\Sigma }}$.

\section{System Model}\label{Sec: System Model}

\subsection{U6G XL-MIMO System Architecture}

\begin{figure}
  \centering
  \includegraphics[scale=0.4]{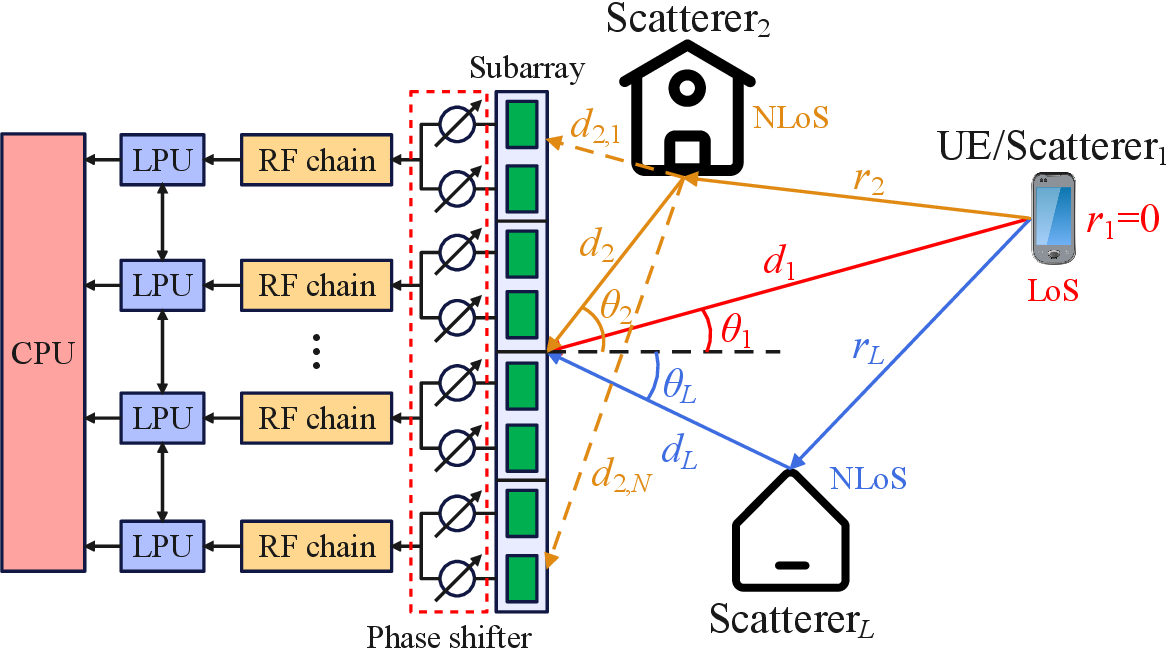}
  \caption{Hybrid XL-MIMO systems with distributed processing architectures.} \label{Fig: system model}
\end{figure}

In this paper, we consider a U6G wideband XL-MIMO system with a distributed hybrid precoding architecture. As shown in Fig.~\ref{Fig: system model}, the BS is equipped with an $N$-antenna uniform linear array (ULA). To reduce the hardware cost of the XL-MIMO systems, the entire array is uniformly divided into $K$ subarrays, whereas each subarray has $N_{\rm s} = \frac{N}{K}$ antennas. Each subarray is connected with a single radio frequency (RF) chain, while an analog phase shifter is deployed for each antenna. Moreover, the distributed processing strategy is considered to decrease the computational overhead of a high-dimensional signal processing process. The received signal of each subarray is processed at the corresponding LPU firstly. Then, the signals processed by the $K$ LPUs are sent to the same CPU.\footnote{As shown in Fig.~\ref{Fig: system model}, the signal processed by an LPU can also be sent to the adjacent LPUs.} Hence, $K$ LPUs can work in parallel or in serial. Consequently, the high-dimensional signal processing process can be divided into several parts of separate low-dimensional processing tasks, and finally combined by the CPU, thus greatly reducing the complexity for XL-MIMO systems with a large number of antennas.

The carrier frequency is denoted as $f_{\rm c}$, and the wavelength is $\lambda _{\rm c} = \frac{c}{f_{\rm c}}$, where $c$ denotes the speed of light. The bandwidth is expressed as $B = M\Delta f$, where $M$ denotes the number of subcarriers, while $\Delta f$ is the frequency spacing between adjacent subcarriers. The frequency of the $m$-th subcarrier can be expressed as ${f_m} = {f_{\rm{c}}} + \delta_{M,m}\Delta f$, where the subcarrier index is $\delta_{M,m} = {m - \frac{{M + 1}}{2}}$, for $m=1,\ldots,M$. The spacing between adjacent antennas is set as $s = \frac{\lambda_{\rm c}}{2}$, while the number of paths in the channel environment is indicated as $L$, which contains a line-of-sight (LoS) path (for $l=1$), and $L-1$ non-LoS (NLoS) paths (for $l = 2,\ldots,L$). Without loss of generality, only a single-antenna UE is considered in this paper. For the multiple-UE scenario, the channels of multiple UEs can be decoupled and individually estimated based on the orthogonal pilot sequences. When the UE transmits an uplink pilot signal to the BS (assume that the transmitted pilot is equal to 1), the received wideband signal of the $k$-th subarray, ${\bf y}_k \in \mathbb{C}^{M \times 1}$, can be denoted as
\begin{equation}\label{Eq: Section2_1}
{\bf y}_{k}^{\rm T} = \sqrt{P}{\bf f}_k^{\rm H}{\bf H}_{k} + {\bf f}_k^{\rm H} {\bf N}_k,
\end{equation}
for $k = 1,\ldots,K$, where $P$ denotes the transmit power; ${\bf H}_{k} \in \mathbb{C}^{N_{\rm s} \times M}$ represents the wideband near-field channel between the UE and the $k$-th subarray; ${\bf f}_k \in \mathbb{C}^{N_{\rm s} \times 1}$ is the analog phase shift vector of the $k$-th subarray, whose each element has constant amplitude and random phase, i.e., the amplitude satisfies $\left|\left[{\bf f}_{k}\right]_m\right| = \frac{1}{\sqrt{N_{\rm s}}}$, while the phase follows a uniform distribution within ${\cal U}\left(0,2\pi\right)$; ${\bf N}_{k} \in \mathbb{C}^{N_{\rm s} \times M}$ denotes the complex additive Gaussian white noise (AWGN) matrix, which satisfies $\left[{\bf N}_{k}\right]_{n,m} \sim {\cal C }{\cal N}\left( 0,\sigma^2 \right)$, for $n = 1,\ldots,N_{\rm s}$, and $m = 1,\ldots,M$. Note that ${\bf z}_k^{\rm T} = {\bf f}_{k}^{\rm H} {\bf N}_k \in \mathbb{C}^{M \times 1}$ is a complex AWGN vector, whose elements have mean zero and variance $\sigma^2$ \cite{near-field_CE2}. Then, the received signal at $K$ subarrays ${\bf Y}_{\rm wn} \in \mathbb{C}^{K \times M}$ can be expressed as
\begin{equation}\label{Eq: Section2_2}
{\bf Y}_{\rm wn} = \sqrt{P}{\bf A}{\bf H}_{\rm wn} + {\bf Z},
\end{equation}
where ${\bf Y}_{\rm wn} = \left[{\bf y}_{1},\cdots,{\bf y}_{K} \right]^{\rm T}$; ${\bf Z} \in \mathbb{C}^{K \times M}$ is a complex AWGN matrix, which satisfies ${\bf Z} = \left[{\bf z}_{1},\ldots,{\bf z}_{K} \right]^{\rm T}$; ${\bf H}_{\rm wn} = \left[{\bf H}_{1}^{\rm T},\ldots,{\bf H}_{K}^{\rm T} \right]^{\rm T}$ denotes the full-dimensional wideband near-field channel; ${\bf A} \in \mathbb{C}^{K \times N}$ is the overall analog phase shift matrix, which can be expressed as
\begin{equation}\label{Eq: Section2_3}
{\bf A} = {\rm blkdiag}\left({\bf f}_{1}^{\rm H},\ldots,{\bf f}_{K}^{\rm H}\right).
\end{equation}

\subsection{Wideband Near-Field Channel Model}

Due to the large array sizes in XL-MIMO systems and wide bandwidths in U6G frequency bands, the near-field and beam squint effects are both non-negligible for U6G XL-MIMO systems. Based on \cite{defineofdualwideband}, the full-dimensional wideband near-field channel ${\bf H}_{\rm wn} \in \mathbb{C}^{N \times M}$ can be denoted as
\begin{equation}\label{Eq: Section2_4}
\begin{aligned}
&{{\bf H}_{\rm wn}} =\\ &\sum\limits_{l=1}^{L} {g_l}e^{j\frac{2\pi}{c}f_{\rm c}\left(r_l+d_l\right)}\left({\bf w}\left(\theta_l,d_l\right){\bf p}^{\rm T}\left(d_l,r_l\right)\right) \odot {\bf Q}_{\rm wn} \left(\theta_l,d_l\right),
\end{aligned}
\end{equation}
where ${g_l} \sim {\cal C}{\cal N}\left(0,1\right)$ denotes the complex gain of the $l$-th path; $\theta_l$ is the sine of the azimuth angle in the $l$-th path as shown in Fig.~\ref{Fig: system model}; $d_l$ and $r_l$ represent the distance from the $l$-th scatterer to the center of the ULA, and the range from the UE to the $l$-th scatterer,\footnote{In the LoS path, the UE is also regarded as a scatterer, such that $r_1 = 0$.} respectively. The antenna-domain near-field steering vector ${\bf w}\left(\theta_l,d_l\right) \in \mathbb{C}^{N \times 1}$ is
\begin{equation}\label{Eq: Section2_5}
{\bf w}\left(\theta_l,d_l\right) = \left[ e^{j\frac{2\pi}{c}f_{\rm c}\Delta d_{1,l}},\cdots,e^{j\frac{2\pi}{c}f_{\rm c}\Delta d_{N,l}}\right]^{\rm T},
\end{equation}
for $l = 1,\ldots,L$, where $\Delta d_{n,l} = d_{n,l} - d_l$ denotes the distance variation from the $l$-th scatterer to the center of the ULA, and to the $n$-th antenna of the ULA. The distance between the $l$-th scatterer and the $n$-th antenna of the BS can be denoted as
\begin{equation}\label{Eq: Section2_6}
{d_{n,l}} = \sqrt{d_{l}^2-2d_l\delta_{N,n}s\theta_l+\delta_{N,n}^2 s^2},
\end{equation}
where the antenna index is $\delta_{N,n} = n - \frac{N+1}{2}$, for $n = 1,\ldots,N$. Note that the distance variation satisfies
\begin{equation}\label{Eq: Section2_7}
{\Delta d_{n,l}} \approx -\delta_{N,n}s\theta_l+\frac{\delta_{N,n}^2s^2\left(1-\theta_l^2\right)}{2d_l},
\end{equation}
due to $d_{n,l} \mathop \approx \limits^{\left( {\rm a} \right)} d_l -\delta_{N,n}s\theta_l + \frac{\delta_{N,n}^2s^2\left(1-\theta_l^2\right)}{2d_l}$, where $\left(\rm a\right)$ is derived by $\sqrt{1+x} \approx 1+\frac{1}{2}x-\frac{1}{8}x^2$ \cite{near-field_CE1}. Meanwhile, the frequency-domain steering vector ${\bf p}\left(d_l,r_l\right) \in \mathbb{C}^{M \times 1}$ is
\begin{equation}\label{Eq: Section2_8}
{\bf p}\left(d_l,r_l\right) = \left[ e^{j\frac{2\pi}{c}\delta_{M,1} \Delta f\left(r_l+d_l\right)},\cdots,e^{j\frac{2\pi}{c}\delta_{M,M} \Delta f\left(r_l+d_l\right)}\right]^{\rm T}.
\end{equation}
The near-field beam squint matrix ${\bf Q}_{\rm wn}\left(\theta_l,d_l\right) \in \mathbb{C}^{N \times M}$ can be expressed according to
\begin{equation}\label{Eq: Section2_9}
\begin{aligned}
&{\bf Q}_{\rm wn}\left(\theta_l,d_l\right) \\= &\left[ {\begin{array}{*{20}{c}}
{e^{j\frac{{2\pi }}{c}\delta_{M,1}\Delta f\Delta {d_{1,l}}}}& \cdots &{e^{j\frac{{2\pi }}{c}\delta_{M,M}\Delta f\Delta {d_{1,l}}}}\\
 \vdots & \ddots & \vdots \\
{e^{j\frac{{2\pi }}{c}\delta_{M,1}\Delta f\Delta {d_{N,l}}}}& \cdots &{e^{j\frac{{2\pi }}{c}\delta_{M,M}\Delta f\Delta {d_{N,l}}}}
\end{array}} \right].
\end{aligned}
\end{equation}

We point out that the entire array is divided into $K \gg 1 $ subarrays under distributed hybrid architectures, whereas each subarray has a smaller aperture than the entire array. Hence, the beam squint within each subarray can be ignored, and only among $K$ subarrays needs to be considered. Thus, the channel of the $k$-th subarray, ${\bf H}_k \in \mathbb{C}^{N_s \times M}$, can be simplified to
\begin{equation}\label{Eq: Section2_10}
{\bf H}_k \approx \sum\limits_{l=1}^L \rho_l \left({\bf w}_k\left(\theta_{l},d_{l}\right){\bf p}^{\rm T}\left(d_{l},r_{l}\right)\right) \odot {\bf Q}_{\rm s}\left(\tilde \theta_{k,l},\tilde d_{k,l}\right),
\end{equation}
where $\rho_l = g_le^{j\frac{2\pi}{c}f_{\rm c}\left(r_l+d_l\right)}$; ${\bf w}_k\left(\theta_l,d_l\right) \in \mathbb{C}^{N_{\rm s} \times 1}$ is the antenna-domain near-field steering vector of the $k$-th subarray; $\tilde \theta_{k,l}$ denotes the sine of the azimuth angle from the $l$-th path observed by the center of the $k$-th subarray, such that
\begin{equation}\label{Eq: Section2_11}
{\tilde \theta}_{k,l} = \frac{d_l\theta_l-\delta_{K,k} N_{\rm s}s}{\sqrt{d_l^2-2\delta_{K,k}d_l N_{\rm s}s\theta_l+\delta_{K,k}^2 N_{\rm s}^2s^2}},
\end{equation}
where $\delta_{K,k} = k - \frac{K+1}{2}$, for $k = 1,\ldots,K$, while $\tilde d_{k,l}$ is the distance from the $l$-th scatterer to the center of the $k$-th subarray, which can be defined according to
\begin{equation}\label{Eq: Section2_12}
\tilde d_{k,l} = \sqrt{d_l^2 - 2 \delta_{K,k} d_l N_{\rm s} s \theta_l + \delta_{K,k}^2 N_{\rm s}^2s^2}.
\end{equation}
Also, ${\bf Q}_{\rm s}\left(\tilde \theta_{k,l},\tilde d_{k,l}\right) \in \mathbb{C}^{N_{\rm s}\times M}$ characterizes the beam quint effect on the $k$-th subarray, which can be expressed as
\begin{equation}\label{Eq: Section2_14}
{\bf Q}_{\rm s} \left(\tilde \theta_{k,l},\tilde d_{k,l}\right) = {\bf p}^{\rm T}\left(\tilde d_{k,l}, -d_l\right)\otimes {\bf 1}_{N_{\rm s}}.
\end{equation}
The full-dimensional channel can be alternatively denoted as
\begin{equation}\label{Eq: Section2_16}
{\bf H}_{\rm wn} \approx \sum\limits_{l=1}^L \rho_l \left( {\bf w} \left(\theta_l, d_l\right) {\bf p}^{\rm T} \left(d_l, r_l\right) \right) \odot {\bf {\tilde Q}} \left(\theta_l, d_l\right),
\end{equation}
where ${\bf {\tilde Q}} \left(\theta_l, d_l\right) \in \mathbb{C}^{N \times M}$ is given by
\begin{equation}\label{Eq: Section2_16_2}
{\bf {\tilde Q}} \left(\theta_l, d_l\right) = \left[{\bf Q}_{\rm s}^{\rm T} \left(\tilde \theta_{1,l},\tilde d_{1,l}\right),\cdots,{\bf Q}_{\rm s}^{\rm T} \left(\tilde \theta_{K,l},\tilde d_{K,l}\right) \right]^{\rm T}.
\end{equation}

Although the beam squint effect within a single subarray can be ignored, it still dominates among different subarrays, which will undermine the polar-domain sparsity of the channel. Thus, conventional near-field channel estimation methods, such as compressed sensing (CS)-based methods, face challenges in wideband channel estimation, whilst a large computational complexity will be required, since those methods rely on full-sampling polar-domain dictionaries. Moreover, distributed hybrid architectures will also pose challenges for channel estimation, due to the small number of RF chains.

Interestingly, in XL-MIMO systems, the near-field channels experiencing beam squint can provide novel degrees of freedom, such as the \textit{parametric symmetry} and spatial-domain row sparsity, which can be explored to simplify the estimation process. Furthermore, the \textit{distributed estimation strategies} could also decrease the computational complexity of wideband near-field channel estimation. These observations will be discussed in the next section.

\section{Distributed Wideband Channel Estimation Based on Parametric Symmetry}\label{Sec: Channel Estimation}

To estimate the wideband near-field channel in a distributed hybrid architecture and under the beam squint effect, the \textit{parametric symmetry} is explored firstly, which can decouple the channel parameters by only linearly combining the delays of different antennas. Based on this, a \textit{distributed estimation algorithm} is proposed, which can greatly reduce the computational complexity through the collaboration between LPUs and the CPU, while requiring only a single pilot.

\subsection{Parametric Symmetry of Wideband Near-Field Channels}

Due to the near-field and beam squint effects, the channel parameters are coupled. However, the delays observed across the array are functions of $\left\{\theta,d,r\right\}$ and the antenna index $\delta_{N,n}$, where $\delta_{N,n}$ and $\delta_{N,n}^2$ are the odd and even symmetries of $n$, respectively. Thus, the partial channel parameters can be decoupled by reasonably combining the delays of different antennas as follows, and can then be readily estimated.

Although the spatial-domain sparsity of wideband near-field channels is broken by the beam squint effect, the delay-domain sparsity is still present, if we observe the received signal from each antenna individually. For simplicity, the single-path scenario is analyzed firstly, and then the multiple-path scenario will be considered in the proposed DPS algorithm. The single-path channel between the UE and the $n$-th antenna of the BS can be expressed as
\begin{equation}\label{Eq: Section3_1}
\left[{\bf H}_{\rm wn}\right]_{n,:} = ge^{j\frac{2\pi}{c}f_{\rm c}\left(r + d +\Delta d_n \right)} {\bf p}^{\rm T}\left(d,r \right) \odot \left[{\bf Q}_{\rm wn}\left(\theta,d\right)\right]_{n,:},
\end{equation}
which can be alternatively denoted according to
\begin{equation}\label{Eq: Section3_2}
\left[{\bf H}_{\rm wn}\right]_{n,:} = ge^{j\frac{2\pi}{c}f_{\rm c}\left(r+d_n\right)} {\bf b}^{\rm T}\left(\tau_n \right),
\end{equation}
where $\tau_n = \frac{\Delta f}{c} \left(r+d_n\right)$, while $d_n = d+\Delta d_n$; ${\bf b}\left(\tau_n \right) = {\bf p}^{\rm T}\left(d,r\right) \odot \left[{\bf Q}_{\rm wn}\left(\theta,d\right)\right]_{n,:}$, for $n = 1,\ldots,N$. The delay-domain steering vector ${\bf b}\left(\tau_n \right) \in \mathbb{C}^{M \times 1}$ is
\begin{equation}\label{Eq: Section3_3}
{\bf b}\left(\tau_n \right) = \left[e^{j 2\pi \delta_{M,1} \tau_{n}},\cdots,e^{j 2\pi \delta_{M,M} \tau_{n}} \right]^{\rm T}.
\end{equation}

In the U6G frequency bands, $r$ and $d_n$ always satisfy $0<\frac{\Delta f}{c}\left(r+d_n\right) < 1$ in the near-field region. Hence, $\tau_{n}$ can be easily estimated by CS-based methods, and, then, based on \eqref{Eq: Section2_7}, the combined value of the range and distance for the $n$-th antenna $\eta \left(n\right) = r + d_n$ can be uniquely calculated as
\begin{equation}\label{Eq: Section3_4}
\eta \left(n\right) = \frac{c}{\Delta f} \tau_{n} = r + d -\delta_{N,n}s\theta + \frac{\delta_{N,n}^2s^2\left(1-\theta^2\right)}{2d}.
\end{equation}
Note that $\delta_{N,n}$ is an odd symmetry of $n$, such that
\begin{equation}\label{Eq: Section3_5}
\delta_{N,n} = -\delta_{N,N-n+1}.
\end{equation}
Meanwhile, $\delta_{N,n}^2$ is an even symmetry of $n$, which can be denoted according to
\begin{equation}\label{Eq: Section3_6}
\delta_{N,n}^2 = \delta_{N,N-n+1}^2.
\end{equation}

\subsubsection{Angle decoupling}

The parameter-dependent function $\eta \left(n\right)$ has an even parametric symmetry w.r.t. $n$, which can be denoted as
\begin{equation}\label{Eq: Section3_7}
\eta \left(N-n+1\right) - \eta \left(n\right) = 2\delta_{N,n}s\theta.
\end{equation}
It is interesting that when the delays of the $n$-th and $\left(N-n+1\right)$-th antennas are obtained, $\theta$ can be simply decoupled by linearly combining them as \eqref{Eq: Section3_7}, and is estimated with low computational complexity.

\subsubsection{Distance decoupling}

Based on the estimated $\theta$, the distance parameter can also be decoupled, since
\begin{equation}\label{Eq: Section3_8}
\delta_{N,\frac{N}{2}+n} - \delta_{N,n} = \frac{N}{2},
\end{equation}
and
\begin{equation}\label{Eq: Section3_9}
\delta_{N,\frac{N}{2}+n}^2 - \delta_{N,n}^2 = N\delta_{N,n} + \frac{N^2}{4},
\end{equation}
for $n = 1,\ldots,\frac{N}{2}$, such that
\begin{equation}\label{Eq: Section3_10}
\begin{aligned}
&\eta \left(\frac{N}{2} + n\right) - \eta \left(n\right) \\= &-\frac{N}{2}s\theta + \frac{s^2\left(1-\theta^2\right)}{2d} \left(N\delta_{N,n}+\frac{N^2}{4}\right).
\end{aligned}
\end{equation}
Hence, when the delays of the $n$-th and $\left(\frac{N}{2} + n\right)$-th antennas are obtained, $d$ can be decoupled by only linearly combining them as \eqref{Eq: Section3_10}, and can then be succinctly estimated from the received signal.

\subsubsection{Range estimation}

Finally, $r$ can also be estimated based on the parametric symmetry of the wideband near-field channels, since the relation between the delays of the $n$-th and $\left(N-n+1\right)$-th antennas can be alternatively expressed as
\begin{equation}\label{Eq: Section3_11}
\eta \left(N-n+1\right) + \eta \left(n\right) = 2r + 2d + \frac{\delta_{N,n}^2 s^2\left(1-\theta^2\right)}{d},
\end{equation}
for $n = 1,\ldots,\frac{N}{2}$. Based on the estimated $\left\{\theta,d\right\}$, and the detected delays of the $n$-th and $\left(N-n+1\right)$-th antennas, $r$ can be easily obtained. We refer to \eqref{Eq: Section3_7}, \eqref{Eq: Section3_10}, and \eqref{Eq: Section3_11} as the parametric symmetry. Subsequently, the wideband near-field channel with a single path can be succinctly estimated based on the parametric symmetry, and the accuracy will be further improved with an increase in the number of antennas, which will be specifically analyzed in Section $\rm \uppercase\expandafter{\romannumeral4}$.

\textbf{\textit{Discussion:}} Note that most existing works \cite{near-field_CE1,near-field_CE2,near-field_CE3,near-field_CE4,near-field_CE5} estimate the spatial parameters depending on a polar-domain dictionary, and require $\frac{N}{2K}$ pilots when equipped with hybrid architectures with fewer RF chains, to scan the whole near-field region. Different from them, and based on the \textit{parametric symmetry}, the spatial parameters can be decoupled and estimated by linearly combining the delays observed from different antennas. Thus, the estimation procedure does not rely on the polar-domain dictionaries, and it entails only \textit{a single pilot}. Therefore, if the parametric symmetry is reasonably utilized in U6G XL-MIMO systems, it can greatly reduce the computational complexity and the number of pilots in wideband near-field channel estimation.

Nevertheless, in real U6G XL-MIMO systems, \textit{hybrid precoding architectures} are usually deployed to reduce the hardware cost, while \textit{multiple paths} generally exist in the U6G frequency bands. Fortunately, the parametric symmetry is still valid in the U6G XL-MIMO systems with hybrid precoding architectures, and can be leveraged to estimate the multiple-path channel effectively.

\subsection{The DPS Algorithm for Channel Estimation}

Since the wideband near-field channel can be modeled as \eqref{Eq: Section2_16}, it still has the parametric symmetry across different subarrays even if a subarray-based hybrid architecture is deployed. Specifically, the delay variations of a path between different subarrays are also dependent on the angle, distance and range parameters. Therefore, the channel parameters of a path can still be decoupled from the delays of this path estimated by different subarrays, to reduce the computational complexity. The received signal of the $k$-th subarray ${\bf y}_{k} \in \mathbb{C}^{M \times 1}$ can be expressed as
\begin{equation}\label{Eq: Section3_12}
{\bf y}_{k} = \sum\limits_{l=1}^{L}\xi_{k,l}{\bf b}\left(\tau_{k,l}\right) + {\bf z}_k,
\end{equation}
where $\xi_{k,l} = g_le^{j\frac{2\pi}{c}f_{\rm c}\left(r_l+d_l\right)}{\bf f}_k^{\rm H}{\bf w}_k\left(\theta_{l},d_l\right)$, while $\tau_{k,l} = \frac{\Delta f}{c}\left(r_l + \tilde d_{k,l}\right)$ denotes the delay of the $l$-th path observed by the $k$-th subarray, for $l = 1,\ldots,L$. Note that if we observe the received signal from each subarray individually, the received signal ${\bf y}_k$ still has a delay-domain sparsity, for $k = 1,\ldots,K$. Specifically, the delay-domain received signal can be denoted by ${\bf \tilde y}_k = {\bf F}_{\tau}^{\rm H} {{\bf y}_k}$, where ${\bf F}_{\tau} \in \mathbb{C}^{M \times M}$ is a discrete Fourier transform (DFT) matrix satisfying ${\bf F}_{\tau} = \frac{1}{\sqrt{M}} \left[ {\bf b} \left( \bar \tau_1 \right), \ldots, {\bf b}\left( \bar \tau_M \right) \right]$; $\bar \tau _m = \frac{2m - 1}{2M}$ is the $m$-th discrete sampling point of $\tau$, where $\left|{\bf b}^{\rm H} \left( \bar \tau_i \right) {\bf b} \left( \bar \tau_j \right)\right| = 0$, for $1 \le i \ne j \le M$. Consequently, if the delay variation between any two paths satisfies $\left|\tau_{k,l_i} - \tau_{k,l_j}\right| \ge \frac{2}{M}$, for $1 \le l_i \ne l_j \le L$, the power of the received signal will be concentrated at only $L$ points in ${\bf \tilde y}_k \in \mathbb{C}^{M \times 1}$, with each point representing the delay-domain signal of a single path. Subsequently, the delays of multiple paths can be estimated from ${\bf \tilde y}_k$ iteratively. In each iteration, a new path will be detected and removed from ${\bf \tilde y}_k$, until all the paths have been detected accurately.

Based on the above discussion, a DPS algorithm applicable to the distributed processing architectures, is proposed to effectively estimate the wideband near-field channel. In each iteration, on the basis of the channel model as in \eqref{Eq: Section2_10}, the delays observed from the center of $K$ subarrays are detected at $K$ LPUs, respectively. To reduce the complexity and improve the accuracy when multiple paths exist, the delays observed from different subarrays that belong to the same path are extrapolated across the LPUs. Then, based on the parametric symmetry, the channel parameters are decoupled and estimated at the CPU by linearly combining the estimates from different LPUs. Finally, the complex gain of a path is calculated by each subarray individually, to reconstruct the full-dimensional wideband near-field channel precisely. The collaboration process between the LPUs and the CPU in a single iteration has been summarized in Fig.~\ref{Fig: flow chart}, and the procedure of the $l$-th iteration can be divided into five steps as follows:

\begin{figure}
  \centering
  \includegraphics[scale=0.4]{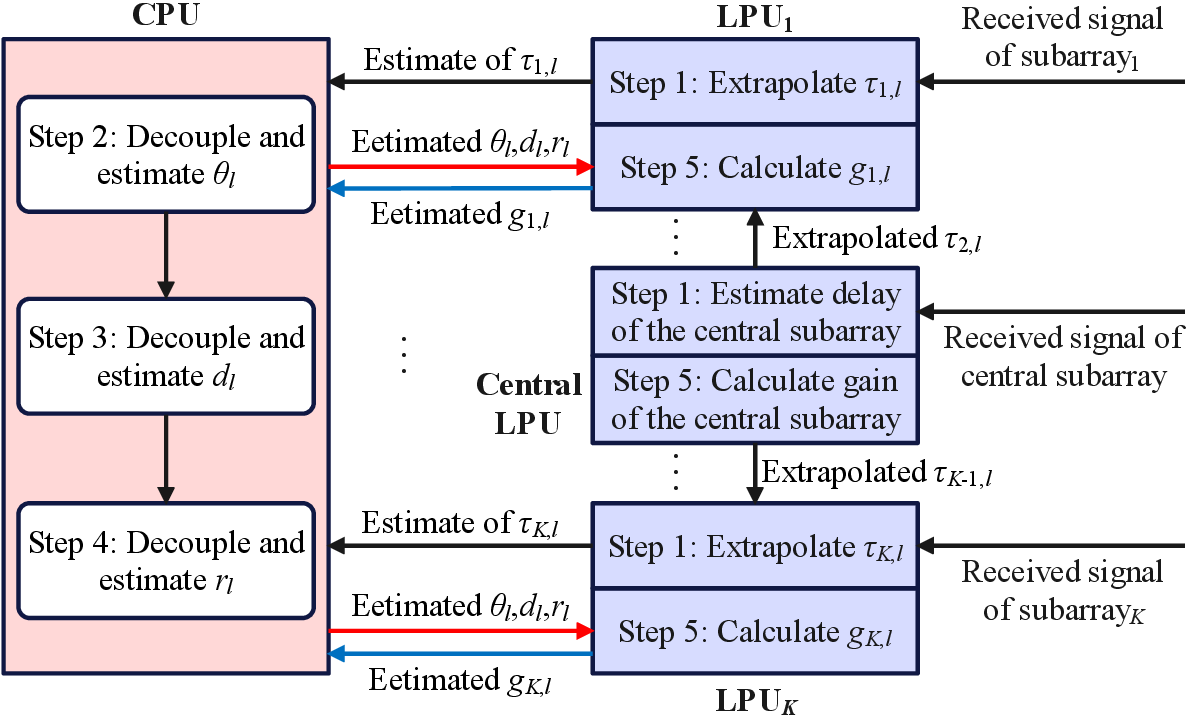}
  \caption{Flow chart of the DPS algorithm between LPUs and the CPU.} \label{Fig: flow chart}
\end{figure}

\subsubsection{Delay extrapolation across LPUs}

Based on the orthogonality between different paths discussed above, the delay of the $k$-th subarray can be easily detected by the maximum likelihood (ML) estimator in the $k$-th LPU, such that
\begin{equation}\label{Eq: Section3_13}
\left\{\hat \xi_{k,l}, \hat \tau_{k,l} \right\} = \mathop {\arg \min }\limits_{\left\{\xi_{k,l}, \tau_{k,l} \right\}} \left\| {\bf y}_k - \xi_{k,l} {\bf b} \left( \tau_{k,l} \right) \right\|^2.
\end{equation}
Based on \cite{ML_estimation}, the formula \eqref{Eq: Section3_13} can be alternatively expressed as
\begin{equation}\label{Eq: Section3_13_2}
\hat \tau_{k,l} = \mathop {\arg \max }\limits_{\tau_{k,l} \in {\cal F}_{\tau}} \frac{\left| {\bf b}^{\rm H} \left( \tau_{k,l} \right) {\bf y}_k \right|^2}{\left\| {\bf b}\left( \tau_{k,l} \right) \right\|^2},
\end{equation}
where the set containing $M$ delay sampling points is given by ${\cal F}_{\tau} = \left\{ \bar \tau_1, \ldots, \bar \tau_M \right\}$ (i.e. $\bar \tau_m$ follows the sampling criteria of the sampling points in the DFT matrix ${\bf F}_{\tau}$, for $m = 1,\ldots,M$). Then, a discrete estimate of $\tau_{k,l}$ can be obtained by \eqref{Eq: Section3_13_2}.

However, given the presence of multiple paths and the large number of subcarriers, it is imperative to acknowledge two fundamental issues. \textbf{Q1}: It is evident that, given the discrete candidate estimate set of $\tau$, there is a high probability that different LPUs will detect delays belonging to different paths in the same iteration. For example, in the $l$-th iteration, the delays of the $l_i$-th and $l_j$-th are detected by the $k_i$-th and $k_j$-th LPUs, respectively, where $1 \le k_i \ne k_i \le K$, and $1 \le l_i \ne l_j \le L$. This will result in challenges for parameter decoupling at the CPU. \textbf{Q2}: Each subarray requires an individual DFT operation, which will consequently lead to a large computational complexity.

The issue \textbf{Q1} can be attributed to the presence of the delay variation in each path between different subarrays, since the index of the detected path in each subarray depends both on the power of each path and the spacing between the real delay value of each path and the closest sampling point \cite{detect_issue}. However, the delay variation of a path between adjacent subarrays is generally significantly smaller than the delay variation between different paths, i.e., $\left| \tau_{k+1,l_i}-\tau_{k,l_i}\right| \ll \left| \tau_{k,l_i} - \tau_{k,l_j} \right|$, for $k = 1,\ldots,K-1$, and $1 \le l_i \ne l_j \le L$. Therefore, the delays of the same path can be extrapolated across different LPUs, which can guarantee that the $K$ LPUs will detect the same path in each iteration. Specifically, each LPU searches only a small range around the estimate of its adjacent LPU, thus greatly reducing the complexity.  

Note that the delay variation of the $l$-th path between two adjacent subarrays can be denoted by
\begin{equation}\label{Eq: Section3_13_3}
\left|\Delta \tau_{k_i, l}\right| = \left| \tau_{k_i + 1, l} - \tau_{k_i,l} \right| \approx \frac{BN_{\rm s}}{2M f_{\rm c}} \left| \theta_l \right| \le \frac{1}{M} \frac{B N_{\rm s}}{2 f_{\rm c}},
\end{equation}
for $l = 1,\ldots,L$, and $1 \le k_i \ne k_j \le K$. Consequently, the delay of the central subarray is estimated through \eqref{Eq: Section3_13_2} firstly, and is reserved as $\hat \tau_{\hat k,l}$, where ${\hat k} = \left\lfloor \frac{K+1}{2} \right\rfloor$ denotes the index of the central subarray. Subsequently, the delays of the $\left({\hat k} + 1\right)$-th and $\left({\hat k} - 1\right)$-th subarrays are extrapolated simultaneously. Based on \eqref{Eq: Section3_13_3}, the delay of the $\left({\hat k} + 1\right)$-th subarray can be extrapolated as
\begin{equation}\label{Eq: Section3_14}
\hat \tau_{\hat k+1,l} = \hat \tau_{\hat k,l} + {\kappa_{\hat k+1}}\frac{1}{M},
\end{equation}
where $\frac{1}{M}$ denotes the delay-domain resolution, while $\kappa_{\hat k+1} \in \left\{-M_{\rm s}, \ldots, M_{\rm s} \right\}$, where $M_{\rm s} = \left\lceil \frac{BN_{\rm s}}{2f_{\rm c}} \right\rceil \ll M$. The extrapolation coefficient $\kappa_{\hat k + 1}$ is given by
\begin{equation}\label{Eq: Section3_15}
\kappa_{\hat k+1,l} = \mathop {\arg \max }\limits_{\kappa \in \left\{-M_{\rm s}, \ldots, M_{\rm s} \right\}}\left|{\bf b}^{\rm H}\left(\hat \tau_{\hat k,l}+\kappa\frac{1}{M}\right) \left[{\bf Y}_{\rm wn} \right]_{\hat k+1,:}^{\rm T}\right|^2.
\end{equation}
Similarly, the delay of the $\left(\hat k-1\right)$-th LPU can also be extrapolated. The delays of the $K$ subarrays are extrapolated across LPUs in a serial pattern, while the estimated delays at the $K$ LPUs are reserved as $\hat \tau_{k,l}$, for $k = 1,\ldots,K$.

Based on this, all delays estimated by the $K$ LPUs are sent to the CPU.\footnote{Note that when the same delay is detected by all the LPUs, the parametric symmetry becomes invalid since the delay variation across the array becomes smaller than the delay resolution. The CPU will then adaptively switch to conventional CS-based algorithms with more pilots and higher complexity.} Subsequently, based on the parametric symmetry between the subarrays, $\left\{\theta_l,d_l,r_l\right\}$ can be decoupled and estimated at the CPU, with low complexity.

\subsubsection{Angle estimation at the CPU}

Similar to \eqref{Eq: Section3_4}, the combined value of $\left\{\theta_l,d_l,r_l\right\}$ observed from the $k$-th subarray $\eta\left(k\right) = \frac{c}{\Delta f} \tau_{k,l}$ can be written as
\begin{equation}\label{Eq: Section3_16}
\eta \left(k\right) =  r_l + d_l - \delta_{K,k}N_{\rm s}s \theta_l + \frac{\delta_{K,k}^2N_{\rm s}^2s^2\left(1-\theta_l^2\right)}{2d_l},
\end{equation}
for $k = 1,\ldots,K$, while the even parametric symmetry of $\eta \left(k\right)$ can be expressed according to
\begin{equation}\label{Eq: Section3_17}
\eta \left(K-k+1\right) - \eta \left(k\right) = 2\delta_{K,k}N_{\rm s}s\theta_l.
\end{equation}
Thus, $\theta_l$ can still be decoupled from the received signal through the linearly combining operation at the CPU as \eqref{Eq: Section3_17}, even with a distributed hybrid architecture. To improve the estimation accuracy, $\frac{K}{2}$ clusters of delays from the $k$-th and $\left(K-k+1\right)$-th LPUs are combined, for $k = 1,\ldots,\frac{K}{2}$. Then, the relation between $\theta_l$ and $\tau_{k,l}$ can be constructed as
\begin{equation}\label{Eq: Section3_18}
2N_{\rm s}s\left[ {\begin{array}{*{20}{c}}
{{\delta _{K,1}}}\\
 \vdots \\
{{\delta _{K,\frac{K}{2}}}}
\end{array}} \right]\theta_l  = \frac{c}{{\Delta f}}\left[ {\begin{array}{*{20}{c}}
{{\tau _{K,l}} - {\tau _{1,l}}}\\
 \vdots \\
{{\tau _{\frac{K}{2} + 1,l}} - {\tau _{\frac{K}{2},l}}}
\end{array}} \right].
\end{equation}
Note that \eqref{Eq: Section3_18} is a linear equation of $\theta_l$, and thus the least square (LS) method can be adopted to estimate $\theta_l$ accurately, with low computational complexity (which is only in direct proportion to the number of subarrays). Thus, based on the estimated delays $\hat \tau_{k,l}$ (for $k = 1,\ldots,K$) in the \textit{delay extrapolation across LPUs} step, $\theta_l$ can be obtained as
\begin{equation}\label{Eq: Section3_19}
\hat \theta_l  = \frac{c}{2N_{\rm s}s\Delta f}{\left[ {\begin{array}{*{20}{c}}
{{\delta _{K,1}}}\\
 \vdots \\
{{\delta _{K,\frac{K}{2}}}}
\end{array}} \right]^\dag }\left[ {\begin{array}{*{20}{c}}
{{\hat \tau_{K,l}} - {\hat \tau_{1,l}}}\\
 \vdots \\
{{\hat \tau _{\frac{K}{2} + 1,l}} - {\hat \tau _{\frac{K}{2},l}}}
\end{array}} \right].
\end{equation}

\begin{algorithm}[t]
\SetAlgoLined
\KwIn{Received signal ${\bf Y}_{\rm wn}$; ${\bf F}_\tau$; ${\bf A}$; $\varsigma$}
\KwOut{Estimated full-dimensional channel ${\bf \hat H}_{\rm wn}$}
\textbf{Initialization:} $l=0$;\\
\While{$\max \left| {\bf F}_{\tau}^{\rm H} {\bf y}_k \right| > \sqrt{\varsigma}$}
{
$l = l + 1$;\\
\textbf{Delay extrapolation across LPUs:} The delay of the central subarray is detected via \eqref{Eq: Section3_13_2}. Then, the delays of other subarrays are extrapolated by \eqref{Eq: Section3_14};\\
\textbf{Angle estimation at the CPU:} By combining the delays estimated at the $k$-th and $\left(K-k+1\right)$-th LPUs, $\theta_l$ is decoupled and estimated as \eqref{Eq: Section3_19};\\
\textbf{Distance estimation at the CPU:} By combining the estimated $\tau_{k,l}$ and $\tau_{k+\frac{K}{2},l}$, for $k = 1,\ldots,\frac{K}{2}$, $d_l$ is decoupled and estimated as \eqref{Eq: Section3_24};\\
\textbf{Range estimation at the CPU:} Based on the estimated $\left\{\theta_l,d_l\right\}$, $\tau_l$ and $\tau_{K-k+1}$ can be linearly combined, while $r_l$ is calculated by \eqref{Eq: Section3_28};\\
\textbf{Gain calculation at LPUs:} The complex gain is calculated at $K$ LPUs by \eqref{Eq: Section3_29} and updated at the CPU. Then, the received signal is updated as \eqref{Eq: Section3_31};\\
}
Reserve estimates $\left\{\hat \rho_{l}, \hat \theta_l,\hat d_l,\hat r_l \right\}$;\\
Reconstruct the wideband near-field channel via \eqref{Eq: Section2_16}.
\caption{The DPS Algorithm}
\end{algorithm}

\subsubsection{Distance estimation at the CPU}

After the \textit{angle estimation at the CPU} step, the estimate of $\theta_l$ is obtained. Then, $\left\{d_l,r_l\right\}$ need to be estimated. Similar to \eqref{Eq: Section3_10}, the delays of the $k$-th and $\left(\frac{K}{2}+k\right)$-th subarrays can also be combined as
\begin{equation}\label{Eq: Section3_20}
\begin{aligned}
&\eta \left(\frac{K}{2} + k\right) - \eta \left(k\right) \\= &-\frac{K}{2}N_{\rm s}s\theta_l + \frac{N_{\rm s}^2s^2\left(1-\theta_l^2\right)}{2d_l}\left(K\delta_{K,k}+\frac{K^2}{4}\right).
\end{aligned}
\end{equation}
Therefore, by combining $\frac{K}{2}$ clusters of delays observed from the $k$-th and $\left(\frac{K}{2}+k\right)$-th subarrays, for $k = 1,\ldots,\frac{K}{2}$, $d_l$ and $\tau_{k,l}$ satisfy
\begin{equation}\label{Eq: Section3_22}
\begin{aligned}
&d_l{\bf v}_{\rm de}\left( \theta_l, \tau_{1,l},\cdots, \tau_{K,l}\right) \\= &\frac{KN_{\rm s}^2s^2\left( {1 - {\theta_l ^2}} \right)}{2}\left[ {\begin{array}{*{20}{c}}
{{\delta_{K,1}} + \frac{K}{4}}\\
 \vdots \\
{{\delta _{K,\frac{K}{2}}} + \frac{{{K}}}{4}}
\end{array}} \right],
\end{aligned}
\end{equation}
where ${\bf v}_{\rm de}\left( \theta_l, \tau_{1,l},\cdots, \tau_{K,l}\right) \in \mathbb{C}^{\frac{K}{2} \times 1}$ can be expressed as
\begin{equation}\label{Eq: Section3_23}
\begin{aligned}
&{\bf v}_{\rm de}\left( \theta_l, \tau_{1,l},\cdots, \tau_{K,l}\right) \\= &\left[ {\begin{array}{*{20}{c}}
{\frac{c}{{\Delta f}}\left( {{\tau _{\frac{K}{2} + 1,l}} - {\tau _{1,l}}} \right) + \frac{K}{2}N_{\rm s}s\theta_l }\\
 \vdots \\
{\frac{c}{{\Delta f}}\left( {\tau _{K,l} - {\tau _{\frac{K}{2},l}}} \right) + \frac{K}{2}N_{\rm s}s\theta_l }
\end{array}} \right].
\end{aligned}
\end{equation}
Then, based on the estimated $\hat \theta_l$ and $\hat \tau_{k,l}$, for $k = 1,\ldots,K$, $d_l$ can be calculated according to
\begin{equation}\label{Eq: Section3_24}
\begin{aligned}
&\hat d_l =\\ &\frac{KN_{\rm s}^2s^2\left( {1 - {\theta ^2}} \right)}{2}{{\bf v}_{\rm de}^\dag\left(\hat \theta_l,\hat \tau_{1,l},\cdots,\hat \tau_{K,l}\right) }\left[ {\begin{array}{*{20}{c}}
{{\delta _{K,1}} + \frac{K}{4}}\\
 \vdots \\
{{\delta _{K,\frac{K}{2}}} + \frac{K}{4}}
\end{array}} \right].
\end{aligned}
\end{equation}

\subsubsection{Range estimation at the CPU}

When $\left\{\theta_l,d_l\right\}$ have been obtained, only $r_l$ remains to be estimated. Based on the odd parametric symmetry of $\eta \left(k\right)$ as \eqref{Eq: Section3_11}, the delays observed from the $k$-th and $\left(K-k+1\right)$-th subarrays can be alternatively combined as
\begin{equation}\label{Eq: Section3_25}
\eta \left(K-k+1\right) + \eta \left(k\right) = 2r_l + 2d_l + \frac{\delta_{K,k}^2N_{\rm s}^2s^2\left(1-\theta_l^2\right)}{d_l}.
\end{equation}
Then, by combining $\frac{K}{2}$ clusters of delays between the $k$-th and $\left(K-k+1\right)$-th subarrays, for $k = 1,\ldots,\frac{K}{2}$, the relation between $r_l$ and $\tau_{k,l}$ can be constructed as
\begin{equation}\label{Eq: Section3_26}
r_l{{\bf{1}}_{\frac{K}{2}}} = {\bf v}_{\rm re} \left(\theta_l,d_l,\tau_{1,l},\cdots,\tau_{K,l}\right),
\end{equation}
where ${\bf v}_{\rm re}\left(\theta_l,d_l,\tau_{1,l},\cdots,\tau_{K,l}\right) \in \mathbb{C}^{\frac{K}{2} \times 1}$ is given by
\begin{equation}\label{Eq: Section3_27}
\left[ {\begin{array}{*{20}{c}}
{\frac{c}{2{\Delta f}}\left( {{ \tau_{K,l}} + { \tau_{1,l}}} \right) - \frac{{\delta _{K,1}^2N_{\rm s}^2{s^2}\left( {1 - {\theta_l ^2}} \right)}}{2d_l} - d_l}\\
 \vdots \\
{\frac{c}{2{\Delta f}}\left( {{ \tau _{{\frac{K}{2}} + 1,l}} + { \tau _{\frac{K}{2},l}}} \right) - \frac{{\delta _{K,\frac{K}{2}}^2N_{\rm s}^2{s^2}\left( {1 - {\theta_l ^2}} \right)}}{2d_l} - d_l}
\end{array}} \right].
\end{equation}
Hence, based on the estimates of $\left\{\hat \theta_l,\hat d_l,\hat \tau_{k,l}\right\}$ in previous steps, for $k = 1,\ldots,K$, $r_l$ can be calculated according to
\begin{equation}\label{Eq: Section3_28}
\hat r_l = \frac{1}{2}{\bf{1}}_{\frac{K}{2}}^\dag {\bf v}_{\rm re} \left(\hat \theta_l,\hat d_l, \hat \tau_{1,l}, \cdots, \hat \tau_{K,l}\right).
\end{equation}

\subsubsection{Gain calculation at LPUs}

To reconstruct the full-dimensional channel at the CPU, the complex gain $\rho_l$ in \eqref{Eq: Section2_16} needs to be estimated. However, given that the signals received by the $K$ subarrays are not transmitted to the CPU, it is necessary to estimate the complex gain $\rho_l$ at the LPUs. Specifically, the estimates $\left\{\hat \theta_l,\hat d_l,\hat r_l\right\}$ obtained at the CPU are sent back to the $K$ LPUs, respectively. Subsequently, at the $k$-th LPU, the complex gain $\rho_l$ can be calculated by
\begin{equation}\label{Eq: Section3_29}
{\hat \rho}_{k,l} = \frac{{\bf v}_{\rm gc}^{\dag}\left(k,\hat \theta_l,\hat d_l,\hat r_l\right)\left[{\bf Y}_{\rm wn}\right]_{k,:}^{\rm T}}{\sqrt{P}\left\|{\bf v}_{\rm gc}\left(k,\hat \theta_l,\hat d_l,\hat r_l\right)\right\|},
\end{equation}
where the coefficient vector ${\bf v}_{\rm gc}\left(k,\hat \theta_l,\hat d_l,\hat r_l\right) \in \mathbb{C}^{M \times 1}$ is
\begin{equation}\label{Eq: Section3_30}
{\bf v}_{\rm gc}^{\rm T}\left(k,\hat \theta_l,\hat d_l,\hat r_l\right) = {\bf f}_{k}^{\rm H}{\bf w}_k\left(\hat \theta_l,\hat d_l\right){\bf p}^{\rm T}\left(\hat {\tilde d}_{k,l} + \hat d_l,\hat r_l - \hat d_l\right),
\end{equation}
for $k = 1,\ldots,K$, while $\left\{\hat{\tilde \theta}_{k,l},\hat{\tilde d}_{k,l}\right\}$ are calculated by \eqref{Eq: Section2_11} and \eqref{Eq: Section2_12}, respectively. Then, the estimated $l$-th path will be removed from the received signals of the $k$-th LPU, and thus the received signal of the $k$-th LPU is updated according to
\begin{equation}\label{Eq: Section3_31}
\begin{aligned}
&\left[{\bf Y}_{\rm wn}\right]_{k,:} = \\ &\left[{\bf Y}_{\rm wn}\right]_{k,:} - {\hat \rho}_{k,l} {\bf{f}}_k^{\rm H}{\bf w}_k\left(\hat \theta_l,\hat d_l \right) {\bf p}^{\rm T} \left( {\hat {\tilde d}}_{k,l} + \hat d_l, \hat r_l - \hat d_l \right).
\end{aligned}
\end{equation}
Finally, the estimates $\hat \rho_{k,l}$ of the $K$ LPUs are sent to the CPU, respectively. To improve the estimation accuracy, the complex gain is updated at the CPU as $\hat \rho_l = \frac{1}{K} \sum \limits_{k=1}^K \hat \rho_{k,l}$. Then, the algorithm moves on to the next iteration.

$\bullet$ \textit{Stopping criterion:} To determine whether all paths have been detected accurately, a constant false alarm rate criterion is adopted \cite{ML_estimation}. The algorithm terminates the iterations when the residual signal received at any LPU satisfies $\max \left| {\bf F}_{\tau}^{\rm H}{\bf y}_k \right| \le \sqrt{\varsigma}$, where $\varsigma = \sigma^2 \ln {M} - \sigma^2 \ln \left(\ln \left( {1 \mathord{\left/
 {\vphantom {1 \left(1 - P_{\rm fa} \right)}} \right.
 \kern-\nulldelimiterspace} \left(1 - P_{\rm fa} \right)} \right)\right)$, while $0 < P_{\rm fa} \ll 1$ is a nominal false alarm rate.

We assume that a total of $\hat L$ paths are detected by the DPS algorithm, and the estimates are reserved at the CPU as $\left\{\hat \rho_l, \hat \theta_l, \hat d_l,\hat r_l \right\}$, for $l = 1,\ldots,\hat L$. Then, the full-dimensional channel is reconstructed by \eqref{Eq: Section2_16}. The details of the proposed DPS algorithm have been summarized in \textbf{Algorithm 1}.

\subsection{Outlook for the Parametric Symmetry}

In real XL-MIMO systems, uniform planar arrays (UPAs) are usually considered, due to the smaller aperture than the ULAs' when equipped with the same number of antennas. The parametric symmetry also exists in UPAs, and the proposed DPS algorithm can be easily modified. Assume that a UPA with $N_{\rm r}$-row and $N_{\rm c}$-column antennas is deployed along the $x$-axis and $y$-axis. The channel between the UE and the $\left(n_{\rm r}, n_{\rm c}\right)$-th antenna of the UPA can be denoted as
\begin{equation}\label{Eq: Section3_33}
\left[{\cal H}_{\rm wn}\right]_{{n_{\rm r},n_{\rm c},:}} = g e^{j\frac{2\pi}{c}f_{\rm c}\left(r+d_{n_{\rm r}, n_{\rm c}}\right)} {\bf b}^{\rm T}\left(\tau_{n_{\rm r},n_{\rm c}}\right),
\end{equation}
for $n_{\rm r} = 1,\ldots,N_{\rm r}$, and $n_{\rm c} = 1,\ldots,N_{\rm c}$, while $d_{n_{\rm r}, n_{\rm c}}$ indicates the distance between the scatterer and the $\left( n_{\rm r}, n_{\rm c} \right)$-th of the UPA. Similar to \eqref{Eq: Section3_13_2}, $\tau_{n_{\rm r},n_{\rm c}}$ can be obtained by the CS-based methods, such that
\begin{equation}\label{Eq: Section3_34}
\begin{aligned}
&\eta\left(n_{\rm r},n_{\rm c}\right) = \frac{c}{\Delta f}\tau_{n_{\rm r},n_{\rm c}} = r + d_{{n_{\rm r},n_{\rm c}}} \mathop \approx \limits^{\left( {\rm a} \right)} r + d - \delta_{N_{\rm r},n_{\rm r}}s\alpha \\ &+ \frac{\delta_{N_{\rm r},n_{\rm r}}^2 s^2 \left(1 - \alpha^2\right)}{2d} - \delta_{N_{\rm c},n_{\rm c}}s\beta + \frac{\delta_{N_{\rm c},n_{\rm c}}^2 s^2 \left(1 - \beta^2\right)}{2d},
\end{aligned}
\end{equation}
where $\left(\rm a\right)$ is derived by $\sqrt{1+x} \approx 1+\frac{1}{2}x-\frac{1}{8}x^2$ \cite{near-field_CE1}; $\alpha$ and $\beta$ denote the angles of the incident signal between the $x$-axis and $y$-axis, respectively. We can find that the delays of the $\left(n_{\rm r}, n_{\rm c}\right)$-th and $\left( N_{\rm r} - n_{\rm r} + 1, n_{\rm c} \right)$-th antennas satisfy
\begin{equation}\label{Eq: Section3_35}
\eta\left(N_{\rm r} - n_{\rm r} + 1,n_{\rm c}\right) - \eta\left(n_{\rm r},n_{\rm c}\right) =  2\delta_{N_{\rm r},n_{\rm r}}s\alpha.
\end{equation}
Similarly, the delays of the $\left(n_{\rm r}, n_{\rm c}\right)$-th and $\left( n_{\rm r}, N_{\rm c} - n_{\rm c} + 1\right)$-th antennas satisfy
\begin{equation}\label{Eq: Section3_36}
\eta\left(n_{\rm r}, N_{\rm c} - n_{\rm c} + 1 \right) - \eta\left(n_{\rm r},n_{\rm c}\right) =  2\delta_{N_{\rm c},n_{\rm c}}s\beta.
\end{equation}
Consequently, based on the parametric symmetry in wideband XL-MIMO systems with UPAs, the angle parameters $\left\{\alpha,\beta\right\}$ can still be decoupled from the received signal and estimated individually. When the estimates of $\left\{\alpha,\beta\right\}$ have both been obtained, the distance parameter can then be decoupled and estimated. When the delays of the $\left(n_{\rm r}, n_{\rm c}\right)$-th and $\left( \frac{N_{\rm r}}{2} - n_{\rm r}, \frac{N_{\rm c}}{2} - n_{\rm c}\right)$-th antennas are obtained, we can get
\begin{equation}\label{Eq: Section3_37}
\begin{aligned}
&\eta\left(\frac{N_{\rm r}}{2} - n_{\rm r}, \frac{N_{\rm c}}{2} - n_{\rm c} \right) - \eta\left(n_{\rm r},n_{\rm c}\right) \\= & -\frac{N_{\rm r}}{2}s\alpha - \frac{N_{\rm c}}{2}s \beta + \frac{s^2}{2d} \left(1-\alpha^2\right) \left(N_{\rm r}\delta_{N_{\rm r},n_{\rm r}}+\frac{N_{\rm r}^2}{4}\right) \\&+ \frac{s^2}{2d} \left(1-\beta^2\right) \left(N_{\rm c}\delta_{N_{\rm c},n_{\rm c}}+\frac{N_{\rm c}^2}{4}\right).
\end{aligned}
\end{equation}
Based on the estimates of $\left\{ \alpha, \beta \right\}$, $d$ can be easily calculated. Subsequently, $r$ can also be acquired based on the parametric symmetry. Consequently, the parametric symmetry still shows great potential in reducing the computational complexity and the number of pilots in channel estimation for U6G XL-MIMO systems with UPAs.

\section{Performance Analysis of DPS Algorithm}\label{Sec: CRLB Analysis}

To evaluate the estimation performances of the proposed DPS algorithm, the computational complexity, the multiple-path resolution, the CRLB of the estimates in hybrid architectures, and the LB of the estimates in the proposed DPS algorithm are analyzed, respectively, and compared to that of the methods without considering the parametric symmetry.

\subsection{Computational Complexity}

The computational complexity of the proposed DPS algorithm is evaluated as follows. In the \textit{delay extrapolation across LPUs} step, since a distributed processing strategy is adopted, the received signal of each subarray is processed individually. Consequently, ${\mathop{\cal O}\nolimits} \left(M^2\right)$ operations are required to estimate the delay at the central LPU, and ${\mathop{\cal O}\nolimits} \left( M_{\rm s} M\right)$ operations are needed to extrapolate the delay at each adjacent LPU. Therefore, the \textit{delay extrapolation across LPUs} step entails ${\mathop{\cal O}\nolimits} \left(M^2 + \left(K-1\right) M_{\rm s} M\right)$ operations in each iteration, and ${\mathop{\cal O}\nolimits} \left(LM^2 + L\left(K-1\right)M_{\rm s}M\right)$ operations in the $L$ iterations. The computational complexity of the \textit{angle estimation at the CPU}, \textit{distance estimation at the CPU}, and \textit{range estimation at the CPU} steps are all ${\mathop{\cal O}\nolimits} \left({K}\right)$ in each iteration, which is proportional to the number of subarrays. Then, a total of ${\mathop{\cal O}\nolimits} \left({LK}\right)$ operations are required in the $L$ iterations. In the \textit{gain calculation at LPUs} step, the LS method requires ${\mathop{\cal O}\nolimits} \left(KM \right)$ operations in each iteration, while the computational complexity of the $L$ iterations can be denoted by ${\mathop{\cal O}\nolimits} \left(LKM \right)$. Consequently, the total computational complexity of the proposed DPS algorithm can be calculated as
\begin{equation}\label{Eq: Section4_1}
{\mathop{\cal O}\nolimits} \left(LM^2 + LKM_{\rm s}M\right),
\end{equation}
where $M \gg 1$, and $K \gg 1$. The computational complexity is dominated by the \textit{delay extrapolation at LPUs} step. However, it has been shown that only a small number of subcarriers are necessary for accurate delay detection, since the estimation accuracy converges when $M$ is large, which has been proved in Section $\rm \uppercase\expandafter{\romannumeral5}$. Consequently, the computational complexity can be further reduced by decreasing the number of subcarriers for channel estimation.

The computational complexity of the proposed DPS algorithm, and that of other wideband near-field channel estimation algorithms have been summarized in Table~\ref{Tab: Computation Complexity}. Since a distributed estimation strategy is adopted in the proposed DPS algorithm, which can divide the high-dimensional signal processing process into several low-dimensional signal processing tasks, it has a lower computational complexity than that of other algorithms. Furthermore, based on the parametric symmetry, the channel parameters can be decoupled by only linearly combining the estimates of LPUs. Thus, different from the methods using two/three-dimensional spatial-domain dictionaries to estimate $\left\{ \theta_l, d_l, r_l \right\}$ simultaneously, the proposed DPS algorithm estimates $\left\{ \theta_l, d_l, r_l \right\}$ by a \textit{non-dictionary} method, for $l = 1,\ldots,L$, and thus the \textit{angle/distance/range estimation at the CPU} step has a much lower complexity that is only in direct proportion to the number of subarrays.

\begin{table}[htbp]
\centering
\caption{Computational complexity analysis}
\label{Tab: Computation Complexity} 
\begin{tabular}{c|c}   
\hline \textbf{Algorithms} & \textbf{Computational Complexity} \\
\hline Method in \cite{wideband_CE2} & ${\mathop{\cal O}\nolimits} \left( LN^2M \right)$ \\
\hline PSIGW \cite{near-field_CE1} & ${\mathop{\cal O}\nolimits} \left(LKN^2M\right)$ \\
\hline BSOMP \cite{wideband_CE3} & ${\mathop{\cal O}\nolimits} \left(L^2KN^2M \right)$ \\
\hline HF-BSPD \cite{wideband_CE6} & ${\mathop{\cal O}\nolimits} \left( LKN^2M \right)$ \\
\hline Proposed DPS & ${\mathop{\cal O}\nolimits} \left( LM^2 + LKM_{\rm s}M \right)$ \\
\hline
\end{tabular}
\end{table}

\subsection{Multiple-Path Resolution}

In conventional dictionary-based wideband near-field channel estimation methods, multiple paths are generally distinguished from the angle domain.\footnote{In the case of two paths with the same incident direction, a significant correlation is observed between them, even within the near-field region. Consequently, it is challenging to effectively distinguish two paths from the distance domain, if they come from the same incident direction.} Consequently, multiple paths cannot be correctly recognized by the dictionary-based estimation methods if the ``angle spacing'' between any two paths is smaller than the angle-domain resolution of a single subarray. The ``angle spacing'' between the $i$-th and $j$-th paths is defined as $\left| \theta_i - \theta_j \right|$, for $1 \le i \ne j \le L$. For example, when the angle variation between the $i$-th and $j$-th paths satisfies $\left| \theta_i - \theta_j \right| < \frac{1}{N_{\rm s}}$, where $\frac{1}{N_{\rm s}}$ is the angle-domain resolution of a single subarray, there is a high probability that two different paths will be incorrectly recognized as the same path. Therefore, in the U6G hybrid XL-MIMO systems with distributed processing architectures, the multiple-path resolution of dictionary-based methods depends mainly on the number of antennas in a single subarray. However, in the context of future 6G wireless systems, it is probable that the BS will be situated in an area devoid of scatterers. Consequently, the majority of the dominant scatterers in the channel between the UE and the BS are usually located on the UE side, which results in a small ``angle spacing'' between multiple paths. As articulated in the preceding discussion, the dictionary-based methods face several challenges in channel estimation. These include high computational complexity and the issue of path discrimination in dense-scattering scenarios.

In contrast to the methods that distinguish between two paths in the spatial domain, the proposed DPS algorithm recognizes each path in the delay domain. Consequently, different paths can still be correctly detected when the ``angle spacing'' between multiple paths is small, even if they have the same angle parameters. However, given the inherent delay resolution constraint (which depends mainly on the number of subcarriers), the proposed DPS algorithm may also encounter the issue of path discrimination if the delay variation between two paths observed by the central subarray is smaller than $\frac{1}{M}$. Thus, it can be concluded that two paths can be detected accurately by the proposed DPS algorithm when
\begin{equation}\label{Eq: Section4_3}
\left| \tau_{\hat k,i} - \tau_{\hat k,j}\right| \ge \frac{1}{M},
\end{equation}
for $1 \le i \ne j \le L$, where $\frac{1}{M}$ represents the delay-domain resolution. Then, based on the relation that $\tau_i = \frac{\Delta f}{c} \left(r_i+d_i\right)$, the distance and range parameters of any two paths observed by the central subarray should satisfy
\begin{equation}\label{Eq: Section4_4}
\left|d_i+r_i -d_j-r_j\right| \ge \frac{\Delta f}{Mc},
\end{equation}
to ensure a correct path discrimination result. Note that $\frac{\Delta f}{Mc} = \frac{B}{M^2c}$, which means that an improvement in multiple-path resolution can be obtained with an increase in $M$ when the bandwidth remains constant. Therefore, although the proposed DPS algorithm has a constraint on the minimum delay variation between different paths, this issue can be mitigated by increasing the number of subcarriers, although $B$ is constant. This conclusion has also been proved in the LB analysis and in Section $\rm\uppercase\expandafter{\romannumeral5}$.

\textbf{\textit{Discussion:}} It is important to note that in dictionary-based estimation methods, the channel parameters are detected from the antenna-domain signal. However, in the context of hybrid XL-MIMO systems, the antenna-domain signal will be compressed due to a limited number of RF chains. Consequently, the accuracy of the estimated channel parameters will be compromised. In contrast, the proposed DPS algorithm detects channel parameters from the delay-domain signal, which remains unchanged by the compressed antenna-domain signal. This method facilitates the improved accuracy of the estimates in hybrid precoding architectures. Moreover, since the delay-domain signal of each subarray is processed individually, it is possible to achieve a high estimation performance even in distributed processing strategies.

\subsection{CRLBs of Estimates in Hybrid Architectures}

\begin{figure}
  \centering
  \includegraphics[scale=0.32]{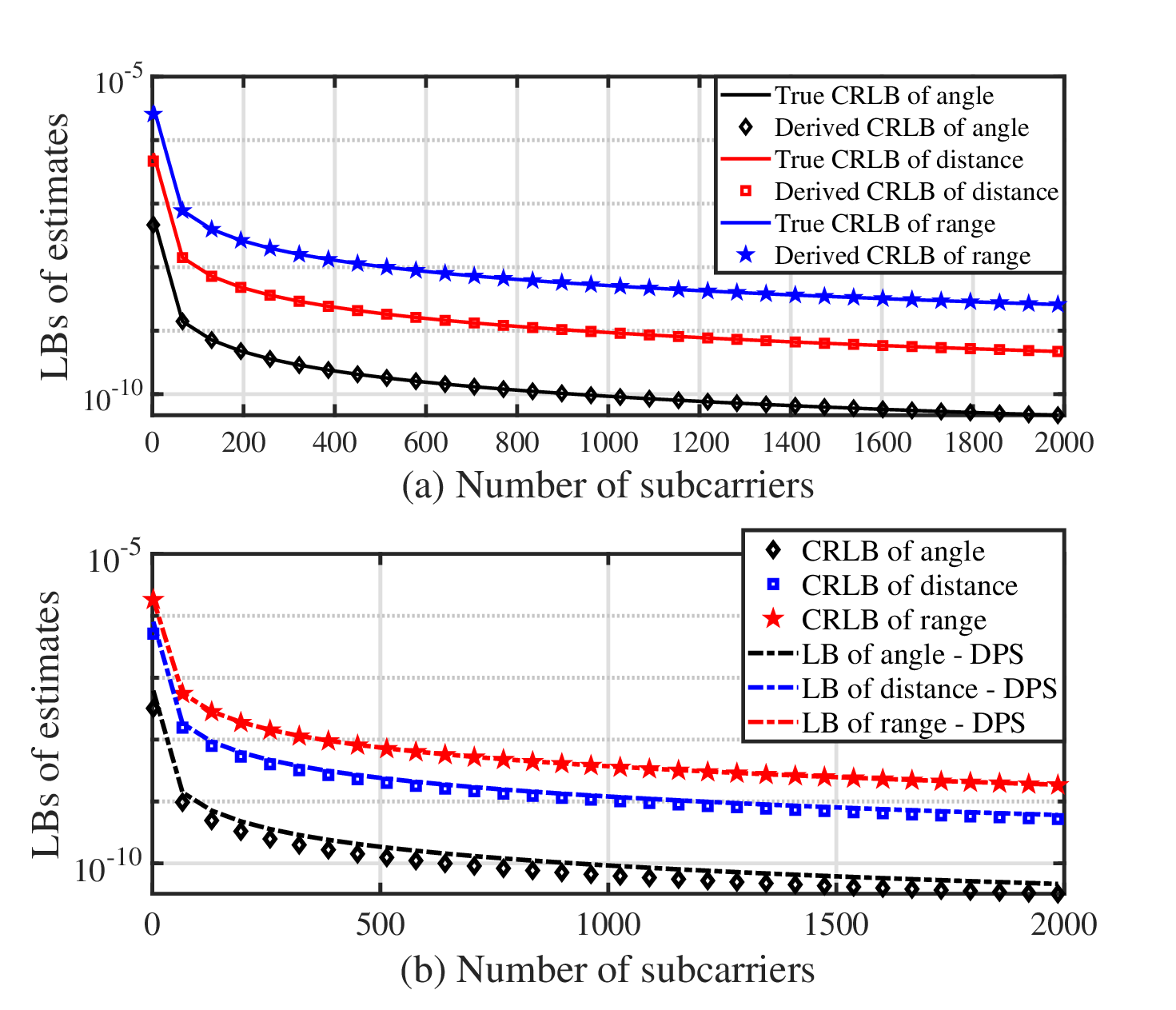}
  \caption{True and closed-form solution of CRLBs, and LBs of distributed processing architectures, where $K = 128$, $\theta = 0.2$, and $d = r = 10$ m.} \label{Fig: CRLB with M}
\end{figure}

To evaluate the influence of the beam squint effect and hybrid architectures on the parameter estimation, the CRLBs of the estimation errors of $\left\{\theta, d, r\right\}$ are derived. Without loss of generality, a single path is considered in this section. Since the received signal is expressed as \eqref{Eq: Section2_2}, the Fisher information matrix (FIM), ${\bf FIM} \in \mathbb{C}^{3 \times 3}$, can then be expressed as
\begin{equation}\label{Eq: Section4_5}
{\bf FIM} = \frac{2P}{\sigma^2} \mathbb{E}\left\{\sum\limits_{k=1}^K \Re \left\{ \left(\frac{\partial {\bf f}_k^{\rm H} {\bf H}_k}{\partial {\bf c}_{\rm pa}}\right)^{\rm H} \frac{\partial {\bf f}_k^{\rm H} {\bf H}_k}{\partial {\bf c}_{\rm pa}}\right\}\right\},
\end{equation}
where ${\bf c}_{\rm pa} = \left[\theta,d,r \right]$. The numerical solution to \eqref{Eq: Section4_5} is challenging to obtain directly, due to the adjustable analog phase shift vector for each subarray, i.e., ${\bf f}_k$, for $k = 1,\ldots,K$. To maximize the power of the received signal, the optimal analog phase shift vectors are designed as ${\bf f}_{k} = \frac{1}{\sqrt{N_{\rm s}}}e^{j\frac{2\pi}{c}f_{\rm c}\left(r + d\right)}{\bf w}_k\left(\theta,d\right)$. Substituting \eqref{Eq: Section2_10} into \eqref{Eq: Section4_5}, such that ${\bf f}_k^{\rm H} {\bf H}_k = g \sqrt{N_{\rm s}} {\bf p}^{\rm T}\left( d - \tilde d_k, r - d \right)$. Thus, the FIM in \eqref{Eq: Section4_5} can be alternatively written as
\begin{equation}\label{Eq: Section4_6}
\begin{aligned}
&{\bf FIM} =\\ &\frac{2PN_{\rm s} \left|g\right|^2}{\sigma^2} \mathbb{E}\left\{\sum\limits_{k=1}^K \Re \left\{ \left(\frac{\partial {\bf u}^{\rm T}\left(k,{\bf c}_{\rm pa}\right)}{\partial {\bf c}_{\rm pa}}\right)^{\rm H} \frac{\partial {\bf u}^{\rm T}\left(k,{\bf c}_{\rm pa}\right)}{\partial {\bf c}_{\rm pa}}\right\}\right\},
\end{aligned}
\end{equation}
where ${\bf u}\left(k,{\bf c}_{\rm pa}\right) \in \mathbb{C}^{M \times 1} = {\bf p} \left( d - \tilde d_k, r - d \right)$ is given by
\begin{equation}\label{Eq: Section4_7}
\begin{aligned}
\left[{\bf u}\left(k,{\bf c}_{\rm pa}\right)\right]_m = e^{j\frac{2\pi}{c}\delta_{M,m}\Delta f\left(r+d+\Delta \tilde d_k\right)},
\end{aligned}
\end{equation}
for $m = 1,\ldots,M$, while $\Delta \tilde d_k = \tilde d_k - d$ denotes the distance variation between the center of the ULA, and the center of the $k$-th subarray. Substituting \eqref{Eq: Section4_7} into \eqref{Eq: Section4_6}, the partial derivative of ${\bf u}\left(k,{\bf c}_{\rm pa}\right)$ can be calculated as
\begin{equation}\label{Eq: Section4_8}
\frac{\partial {\bf u}\left(k,{\bf c}_{\rm pa}\right)}{\partial x} = j\frac{2\pi}{c}\rho_{x,n} \left[ \delta_{M,1}, \cdots, \delta_{M,M} \right]^{\rm T} \Delta f \odot {\bf u}\left(k,{\bf c}_{\rm pa}\right),
\end{equation}
where $x \in {\bf c}_{\rm pa}$, while
\begin{equation}\label{Eq: Section4_9}
\left\{ {\begin{array}{*{20}{l}}
{\rho_{\theta,k} = -\delta_{K,k}N_{\rm s}s-\frac{\delta_{K,k}^2{N_{\rm s}^2}s^2\theta}{d}}\\
{\rho_{d,k} = 1-\frac{\delta_{K,k}^2N_{\rm s}^2s^2\left(1-\theta^2\right)}{2d^2}}\\
{\rho_{r,k} = 1}.
\end{array}} \right.
\end{equation}
Then, we have
\begin{equation}\label{Eq: Section4_10}
{\left( {\frac{\partial {\bf u}\left( {k,{\bf c}_{\rm pa} } \right)}{\partial x}} \right)^{\rm{H}}}\frac{\partial {\bf u}\left( {k,{\bf c}_{\rm pa}} \right)}{\partial y} = \frac{4{\pi ^2}}{c^2} \rho_{x,k} \rho_{y,k} \sum\limits_{m = 1}^M {\delta_{M,m}^2} \Delta f^2,
\end{equation}
where $\left\{x,y\right\} \in {\bf c}_{\rm pa}$, and
\begin{equation}\label{Eq: Section4_11}
\sum\limits_{m = 1}^{M}\delta_{M,m}^2 \Delta f^2 = \frac{M\left(M^2-1\right)}{12}\Delta f^2.
\end{equation}
Hence, the FIM can be alternatively written as
\begin{equation}\label{Eq: Section4_12}
{\bf FIM} = \frac{2\pi^2P\left|g\right|^2 N_{\rm s} M\left(M^2-1\right)\Delta f^2}{3c^2\sigma^2}{\bf U}_{\rm dh},
\end{equation}
where ${\bf U}_{\rm dh} \in \mathbb{C}^{3 \times 3}$ can be expressed according to
\begin{equation}\label{Eq: Section4_13}
{\bf U}_{\rm dh} = \left[ {\begin{array}{*{20}{c}}
{\sum\limits_{k = 1}^K {\rho _{\theta ,k}^2} }&{\sum\limits_{k = 1}^K {{\rho _{\theta ,k}}{\rho _{d,k}}} }&{\sum\limits_{k = 1}^K {{\rho _{\theta ,k}}{\rho _{r,k}}} }\\
{\sum\limits_{k = 1}^K {{\rho _{d,k}}{\rho _{\theta ,k}}} }&{\sum\limits_{k = 1}^K {\rho _{d,k}^2} }&{\sum\limits_{k = 1}^K {{\rho _{d,k}}{\rho _{r,k}}} }\\
{\sum\limits_{k = 1}^K {{\rho _{r,k}}{\rho _{\theta,k}}} }&{\sum\limits_{k = 1}^K {{\rho_{r,k}}{\rho _{d,k}}} }&{\sum\limits_{k = 1}^K {\rho _{r,k}^2} }
\end{array}} \right].
\end{equation}
Subsequently, the CRLB matrix of $\left\{ \theta, d, r\right\}$ can be calculated by ${\bf CB}_{\rm dh} \in \mathbb{C}^{3 \times 3} = {\bf FIM}^{\rm -1}$, such that
\begin{equation}\label{Eq: Section4_14}
{\bf CB}_{\rm dh} = \left(\frac{c\sigma}{\pi \left| g \right|\sqrt{P}}\right)^2\frac{3}{2N_{\rm s}M\left(M^2-1\right)\Delta f^2}{\bf U}_{\rm dh}^{-1},
\end{equation}
while the CRLBs of the estimated channel parameters $\left\{\theta,d,r\right\}$ can be respectively expressed as
\begin{equation}\label{Eq: Section4_15}
\left\{\theta_{\rm CB},d_{\rm CB},r_{\rm CB}\right\} = \left\{\left[{\bf CB}_{\rm dh}\right]_{1,1},\left[{\bf CB}_{\rm dh}\right]_{2,2},\left[{\bf CB}_{\rm dh}\right]_{3,3}\right\}.
\end{equation}

\textbf{\textit{Proposition 1:}} In wideband XL-MIMO systems with hybrid architectures, the closed-form solution of CRLBs for $\left\{\theta,d,r\right\}$ can be expressed according to
\begin{equation}\label{Eq: Section4_16}
\left\{ {\begin{array}{*{20}{l}}
{\theta_{\rm CB} \approx {\frac{c^2\sigma^2}{\pi^2 P \left| g \right|^2}} \frac{{36\left( {1 - {\theta ^2}} \right)}}{{{K^3}N_{\rm s}^3 M^3\Delta f^2{s^2}\left( {2 + 7{\theta ^2}} \right)}}}\\
{d_{\rm CB} \approx {\frac{c^2\sigma^2}{\pi^2 P \left| g \right|^2}} \frac{{144{d^2}\left( {15{d^2} + {s^2}{\theta ^2}{N_{\rm s}^2K^2}} \right)}}{{{K^5}N_{\rm s}^5 M^3\Delta f^2 {s^4}\left( {1 - {\theta ^2}} \right)\left( {2 + 7{\theta ^2}} \right)}}}\\
{r_{\rm CB} \approx {\frac{c^2\sigma^2}{\pi^2 P \left| g \right|^2}} \frac{{8640{d^4} + {s^2}{d^2}{N^2}\left( {1296{\theta ^2} - 720} \right) + 27{s^4}{{\left( {1 - {\theta ^2}} \right)}^2}{N^4}}}{{4{K^5}N_{\rm s}^5 M^3 \Delta f^2{s^4}\left( {1 - {\theta ^2}} \right)\left( {2 + 7{\theta ^2}} \right)}}}.
\end{array}} \right.
\end{equation}
The proof is provided in Appendix A.

When the incident signal is vertical to the normal direction of the array, i.e., $\left|\theta \right| \to 1$, then, $\theta_{\rm CB} \to 0$, which means that $\theta$ can be accurately estimated even with the noise and beam squint effect. However, $d_{\rm CB} \to \infty$ and $r_{\rm CB} \to \infty$, when $\left|\theta \right| \to 1$, because the near-field spherical wave model degrades into the far-field planar wave model, in which case the channel is independent of $d$ and $r$, and thus $\left\{d,r\right\}$ cannot be detected correctly from the received signal. In Fig.~\ref{Fig: CRLB with M}(a), the true and derived closed-form solutions of the CRLBs for $\left\{\theta,d,r\right\}$ are simulated, where the system conditions are set as in Section $\rm \uppercase\expandafter{\romannumeral5}$. It is evident that the derived closed-form solution \eqref{Eq: Section4_16} exhibits a high degree of proximity to the actual result. Consequently, the closed-form solutions of CRLB derived in \eqref{Eq: Section4_16} can be used to analyze the error lower bound of the estimates, and evaluate the influence of the system settings on channel estimation in wideband XL-MIMO systems with hybrid architectures.

\subsection{LBs of Estimates in the Proposed Algorithm}

It is imperative to note that a distributed processing strategy is designed to reduce the computational complexity, in which case each LPU can only process the signal received by a single subarray. However, the CRLBs of the estimates derived in Section $\rm \uppercase\expandafter{\romannumeral4}$.C do not consider this constraint, and, hence, cannot be used directly to evaluate the proposed DPS algorithm. In contrast to the CRLBs of the estimates, which assume that the CPU can obtain the received signals of all the subarrays, a key issue that needs to be addressed is whether the channel parameters can still be estimated accurately when a distributed processing strategy is adopted. Consequently, the LBs of the estimation errors in the distributed processing strategies are then derived, to evaluate the estimation performance of the proposed DPS algorithm. Different from the methods to detect $\left\{\theta,d,r\right\}$ simultaneously, only the delay is estimated at each LPU. Therefore, the FIM of the $k$-th LPU can be expressed according to
\begin{equation}\label{Eq: Section4_17}
\begin{aligned}
&{\rm FIM}_{{\rm DPS},k} =\\ &\frac{2PN_{\rm s} \left|g\right|^2}{\sigma^2} \mathbb{E}\left\{ \Re \left\{ \left(\frac{\partial {\bf b} \left(\tau_k \right)}{\partial \tau_k}\right)^{\rm H} \frac{\partial {\bf b} \left(\tau_k \right)}{\partial \tau_k}\right\}\right\}.
\end{aligned}
\end{equation}
Due to $\frac{\partial {\bf b}\left(\tau_k \right)}{\partial \tau_k} = j2\pi \left[\delta_{M,1},\cdots,\delta_{M,M}\right]^{\rm T} \odot {\bf b}\left(\tau_k \right)$, we can get
\begin{equation}\label{Eq: Section4_18}
\begin{aligned}
\left(\frac{\partial {\bf b} \left(\tau_k \right)}{\partial \tau_k}\right)^{\rm H} \frac{\partial {\bf b} \left(\tau_k \right)}{\partial \tau_k} = 4\pi^2 \sum\limits_{m = 1}^M\delta_{M,m}^2,
\end{aligned}
\end{equation}
for $k = 1,\ldots,K$, while the FIM can then be rewritten as
\begin{equation}\label{Eq: Section4_19}
{\rm FIM}_{{\rm DPS},k} = \frac{2\pi^2PN_{\rm s} \left|g\right|^2 M\left(M^2-1\right)}{3\sigma^2}.
\end{equation}
Since the FIM is independent of the index of the subarray $k$, the delays estimated by $K$ LPUs have the same CRLB, which can be calculated according to
\begin{equation}\label{Eq: Section4_20}
{\tau}_{\rm CB} = \frac{3\sigma^2}{2\pi^2PN_{\rm s} \left|g\right|^2 M\left(M^2-1\right)}.
\end{equation}

Therefore, we assume that the delay estimation errors are i.i.d. random variables, satisfying $\Delta \tau_k = {\hat \tau_k}-{\tau_k} \sim {\cal C}{\cal N} \left(0,{\tau}_{\rm CB}\right)$, for $k = 1,\ldots,K$, and $\Delta \tau_{i,j} = \left({\hat \tau_i}-{\tau_i}\right) - \left({\hat \tau_j}-{\tau_j}\right) \sim {\cal C}{\cal N} \left(0,2{\tau}_{\rm CB}\right)$, for $1 \le i\ne j \le K$. At the CPU, the estimation errors of $\left\{ \theta,d,r \right\}$ by \eqref{Eq: Section3_19}, \eqref{Eq: Section3_24}, and \eqref{Eq: Section3_28}, depend mainly on $\tau_{\rm CB}$. For ease of expression, we take the angle estimation procedure as an example. Based on \eqref{Eq: Section3_18}, we can get
\begin{equation}\label{Eq: Section4_20_2}
\begin{aligned}
& 4N_{\rm s}^2 s^2 \left\| \left[ \delta_{K,1}, \cdots, \delta_{K,\frac{K}{2}} \right] \right\|^2 \mathbb{E} \left\{ \left(\hat \theta - \theta \right)^2\right\} \\= &\frac{c^2}{\Delta f^2} \mathbb{E} \left\{ \left\|\left[ \Delta \tau_{1,K},\cdots,\Delta \tau_{\frac{K}{2},\frac{K}{2}+1} \right]\right\|^2\right\}.
\end{aligned}
\end{equation}
Then, the LB of $\theta$ is obtained as
\begin{equation}\label{Eq: Section4_21}
\begin{aligned}
\theta_{\rm LB} &= \mathbb{E}\left\{\left( \hat \theta - \theta \right)^2 \right\}= \frac{c^2\tau_{\rm CB}}{2N_{\rm s}^2s^2\Delta f^2} \left( \sum\limits_{k = 1}^{\frac{K}{2}} \delta_{K,k}^2\right)^{-1}  \\ &= \frac{c^2\sigma^2}{\pi^2P\left|g\right|^2} \frac{18}{K\left(K^2 -1 \right) N_{\rm s}^3M\left(M^2-1\right)\Delta f^2 s^2},
\end{aligned}
\end{equation}
where $\sum\limits_{k = 1}^{\frac{K}{2}} \delta_{K,k}^2 = \frac{K\left(K^2-1\right)}{24}$. Similarly, based on \eqref{Eq: Section3_22} and \eqref{Eq: Section3_26}, the LBs of $\frac{1}{d}$ and $r$ can be respectively calculated as
\begin{equation}\label{Eq: Section4_22}
\left\{ {\begin{array}{*{20}{l}}
{{\frac{1}{d}}_{\rm LB} = \frac{{{c^2}{\sigma ^2}}}{{{\pi ^2}P{{\left| g \right|}^2}}}\frac{{1152}}{{{K^3}\left( {{K^2} - 4} \right)N_{\rm{s}}^5M\left( {{M^2} - 1} \right)\Delta {f^2}{s^4}{{\left( {1 - {\theta _2}} \right)}^2}}}}\\
{{r_{\rm LB}} = \frac{{{c^2}{\sigma ^2}}}{{{\pi ^2}P{{\left| g \right|}^2}}}\frac{3}{{2K{N_{\rm{s}}}M\left( {{M^2} - 1} \right)\Delta {f^2}}}}.
\end{array}} \right.
\end{equation}
Note that since ${\frac{1}{d}}_{\rm LB} = \mathbb{E} \left\{\left(\frac{1}{\hat d} - \frac{1}{d}\right)^2 \right\} = \mathbb{E} \left\{\frac{\left( \hat d- d \right)^2}{\hat d^2d^2}\right\} \approx \frac{1}{d^4}\mathbb{E}\left\{\left(\hat d- d\right)^2\right\}$, where $d \gg  \left|\hat d-d\right|$, such that the LB of $d$ is then approximately expressed as
\begin{equation}\label{Eq: Section4_24}
\begin{aligned}
&d_{\rm LB} \approx d^4 {\frac{1}{d}}_{\rm LB} = \\ &\frac{{{c^2}{\sigma ^2}}}{{{\pi ^2}P{{\left| g \right|}^2}}}\frac{{1152d^4}}{{K^3}\left(K^2-4\right)N_{\rm{s}}^5M\left( {{M^2} - 1} \right)\Delta {f^2}{s^4}\left( {1 - {\theta ^2}} \right)^2}.
\end{aligned}
\end{equation}

Note that the LBs of $\left\{\theta,d,r\right\}$ in distributed processing architectures $\left\{\theta_{\rm LB},d_{\rm LB},r_{\rm LB}\right\}$ and the CRLBs $\left\{\theta_{\rm CB},d_{\rm CB},r_{\rm CB}\right\}$ are both inversely proportional to ${\cal O}\left(N^3 M^3\right)$, ${\cal O}\left(N^5M^3\right)$, and ${\cal O}\left(N M^3\right)$, respectively, when $\theta \to 0$, where $N = KN_{\rm s}$. When $\theta = 0$, we can get $\theta_{\rm LB} = \theta_{\rm CB}$. As shown in Fig.~\ref{Fig: CRLB with M}(b), the LBs of the estimates match closely with the CRLBs, indicating that the proposed DPS algorithm maintains a superior estimation capability, even when the computational complexity and number of pilots are reduced in distributed processing architectures. The estimation capability is anticipated to be further enhanced by an increase in the number of subcarriers and antennas. As the values of $N$ and the bandwidth approach infinity, the LBs of $\left\{\theta,d,r\right\}$ tend to $0$. Furthermore, since $B = M \Delta f$, the LBs of $\left\{ \theta, d, r \right\}$ are both inversely proportional to $M$ when $B$ is constant. Consequently, increasing $M$ under a bandwidth constraint is an effective approach to enhance the estimation capability of the DPS algorithm.

\subsection{Robustness Analysis under Hardware Impairments}

In practical U6G XL-MIMO systems, the issues of hardware impairments may arise, which can lead to adverse effects on the received signals across different LPUs. In order to evaluate the robustness of the proposed DPS algorithm in the presence of hardware impairments, the effects of clock offsets, gain mismatches and antenna misalignments on delay estimation have been discussed. Without loss of generality, a single-path scenario is considered.

\textbf{\textit{1) Clock offsets:}} In instances where clock offsets exist across different LPUs, this will result in an additional phase shift for the received signal of each LPU and in each subcarrier. Assume that compared to a reference LPU, the $k$-th LPU has a clock offset ${\cal T}_k$, for $k = 1,\ldots,K$. Then, the received signal at the $k$-th LPU is given by
\begin{equation}\label{Eq: SectionR1_2_1}
{\bf y}_k = \xi_k e^{j 2\pi f_{\rm c} {\cal T}_k} {\bf b} \left( \tau_k + {\cal T}_k \right) + {\bf z}_{k}.
\end{equation}
Consequently, the clock offsets across LPUs will result in a variation between the estimated and true delays at the $k$-th LPU as $\Delta \tau_k = {\cal T}_k$. However, it is important to note that the estimates of delay are discrete in the proposed DPS algorithm, which have been quantified as a set of sampling points ${\frac{2m - 1}{2M}}$, for $m = 1,\ldots,M$. The delay at each LPU will be estimated as the closest sampling point, which means that a delay error less than $\frac{1}{M}$ can be permitted, since the spacing between adjacent sampling points is $\frac{2}{M}$ (if the delay error is larger than $\frac{1}{M}$, an incorrect sampling point will be identified). It can thus be concluded that, despite of the occurrence of clock offsets, the proposed DPS algorithm is capable of estimating the spatial parameters by means of parametric symmetry, on the condition that the maximum clock offset satisfies $\max_k {\cal T}_k \le \frac{1}{M}$ (i.e., the decoupling of spatial parameters depends mainly on the delay variation trend along the entire array rather than the exact value of each delay).

\textbf{\textit{2) Gain mismatches and antenna misalignments:}} In instances where gain mismatches and antenna misalignments exist across different LPUs, the power of the received signal at each LPU will be compromised. Assume that the gain factor in each subarray is denoted as $0 < {\cal P}_k \le 1$, for $k = 1,\ldots,K$. The received signal at the $k$-th LPU is given by
\begin{equation}\label{Eq: SectionR1_2_2}
{\bf y}_k = {\cal P}_k \xi_k {\bf b} \left( \tau_k \right) + {\bf z}_{k}.
\end{equation}
Note that the received signal of the $k$-th LPU can be alternatively written as ${\bf y}_k = \xi'_k {\bf b} \left( \tau_k \right) + {\bf z}_{k}$, where $\xi'_k = {\cal P}_k \xi_k$ is the equivalent complex gain of the channel in the $k$-th subarray. Due to the existence of gain mismatches and antenna misalignments, the equivalent gain observed by each LPU will be changed. However, it is evident that, given the individual processing of the received signal within each LPU, the issue of gain mismatches and antenna misalignments will not exert an influence on the extrapolation of delays across LPUs. This phenomenon can be attributed to the fact that the delay extrapolation process does not rely on the value of the equivalent complex gain at each LPU. Furthermore, to accurately estimate the channel by the proposed DPS algorithm, the complex gain is estimated at each LPU individually, which can eliminate the effects of gain mismatches and antenna misalignments in channel reconstruction.

Consequently, the LPUs retain the capacity to attain strong delay estimation performance in scenarios where hardware impairment is not pronounced, thereby verifying the robustness of the proposed DPS algorithm.

\section{Simulation Results}\label{Sec: Simulation Results}

In this section, the normalized mean square error (NMSE) of the proposed DPS algorithm is evaluated, and compared to that of other algorithms. The NMSE is defined as
\begin{equation}\label{Eq: Section5_1}
{\rm{NMSE}} = \frac{{{{\left\| {{\rm{vec}}\left( {{\bf \hat H}_{\rm wn} - {\bf H}_{\rm wn}} \right)} \right\|}^2}}}{{{{\left\| {{\rm{vec}}\left( {\bf H}_{\rm wn} \right)} \right\|}^2}}},
\end{equation}
where ${\bf H}_{\rm wn}$ and ${\bf \hat H}_{\rm wn}$ denote the real and estimated channels, respectively. The signal-to-noise (SNR) is defined as
\begin{equation}\label{Eq: Section5_2}
{\rm{SNR}} = 10{\rm log}_{10}\left(\frac{P\left\|{\rm vec}\left({\bf A}{\bf H}_{\rm wn}\right)\right\|^2}{KM\sigma^2}\right).
\end{equation}

\begin{figure}
  \centering
  \includegraphics[scale=0.3]{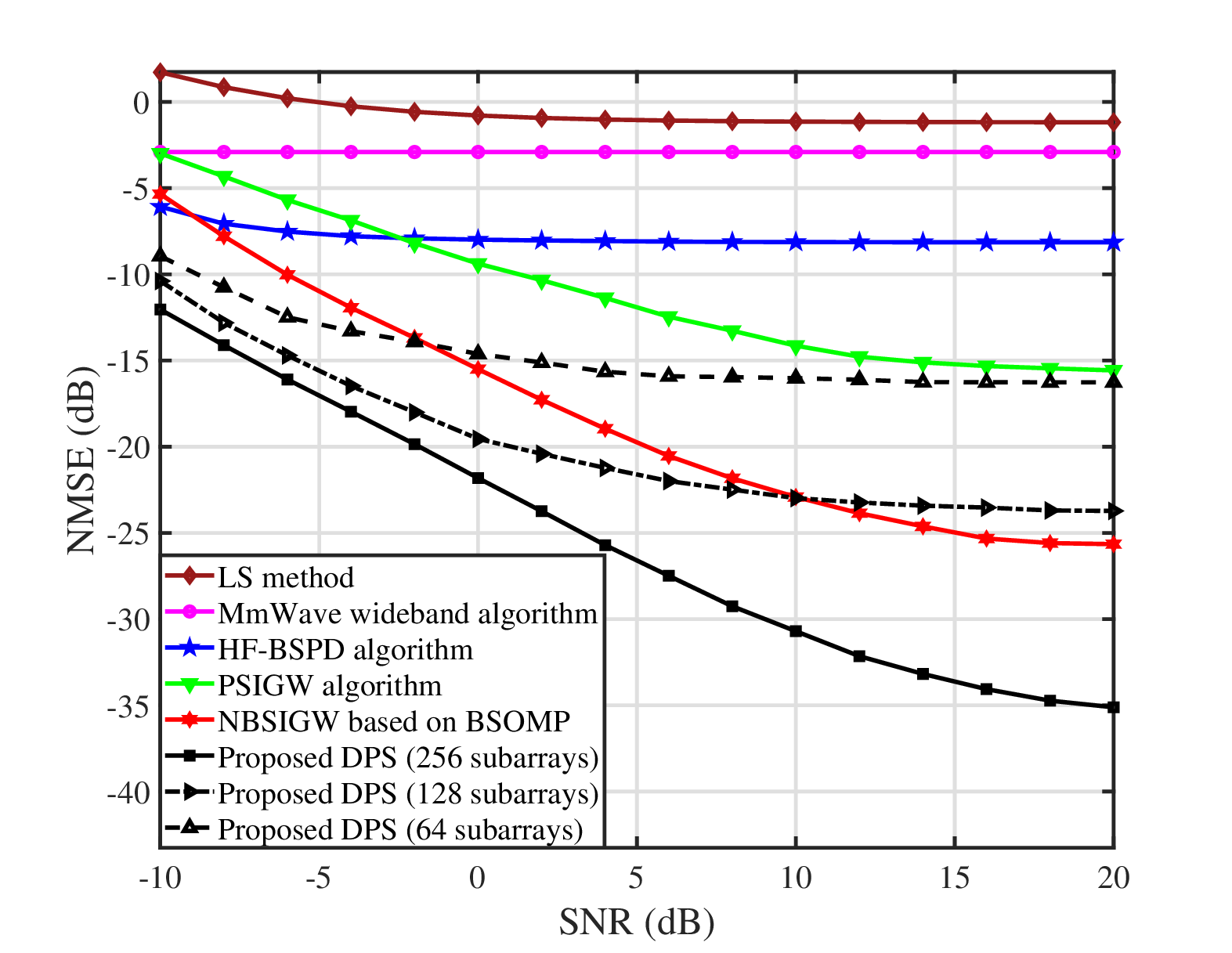}
  \caption{NMSE of the reconstructed wideband near-field channel.} \label{Fig: NMSE against SNR}
\end{figure}

The system conditions of the U6G wideband XL-MIMO system are set as follows: The carrier frequency is ${f_c} = 7$ GHz; the bandwidth is $B = 600$ MHz, while the number of subcarriers is $M = 1024$. The number of antennas is ${N} = 1024$, and the number of subarrays is $K = 64/128/256$; the number of paths is set as $L = 4$; $P_{\rm fa} = 10^{-3}$; $\left\{d,r\right\}$ and $\theta$ follow the uniform distribution within ${\cal U}\left( {10,20} \right)$ m and ${\cal U}\left( {-1, 1} \right)$, respectively. In fairness, an uplink pilot and $K = 256$ are set for other estimation algorithms.

The NMSE of the reconstructed channel against the SNR is evaluated firstly. As shown in Fig.~\ref{Fig: NMSE against SNR}, the limited number of RF chains present in distributed hybrid architectures renders the LS method ineffective for channel estimation. The mmWave wideband algorithm in \cite {wideband_CE2} requires that the LoS path dominates the channel components and has compromised NMSE performance in multiple-path scenarios. The HF-BSPD algorithm in \cite{wideband_CE6} and the PSIGW algorithm in \cite{near-field_CE1} also have compromised NMSE performances, due to the deployment of distributed hybrid architectures. In contrast, the NBSIGW algorithm based on BSOMP in \cite{wideband_CE3} has an improved estimation accuracy, since a wideband dictionary is constructed that incorporates frequency-dependent sparse support under the beam squint effect. However, this approach results in high computational complexity, due to the dictionary scanning operation. Compared to other algorithms, the proposed DPS algorithm yields a lower NMSE, while requiring much reduced computational complexity. It should be noted that the proposed DPS algorithm continues to perform better at low SNRs, while the NMSE performance deteriorates as $K$ decreases.

\begin{figure}
  \centering
  \includegraphics[scale=0.3]{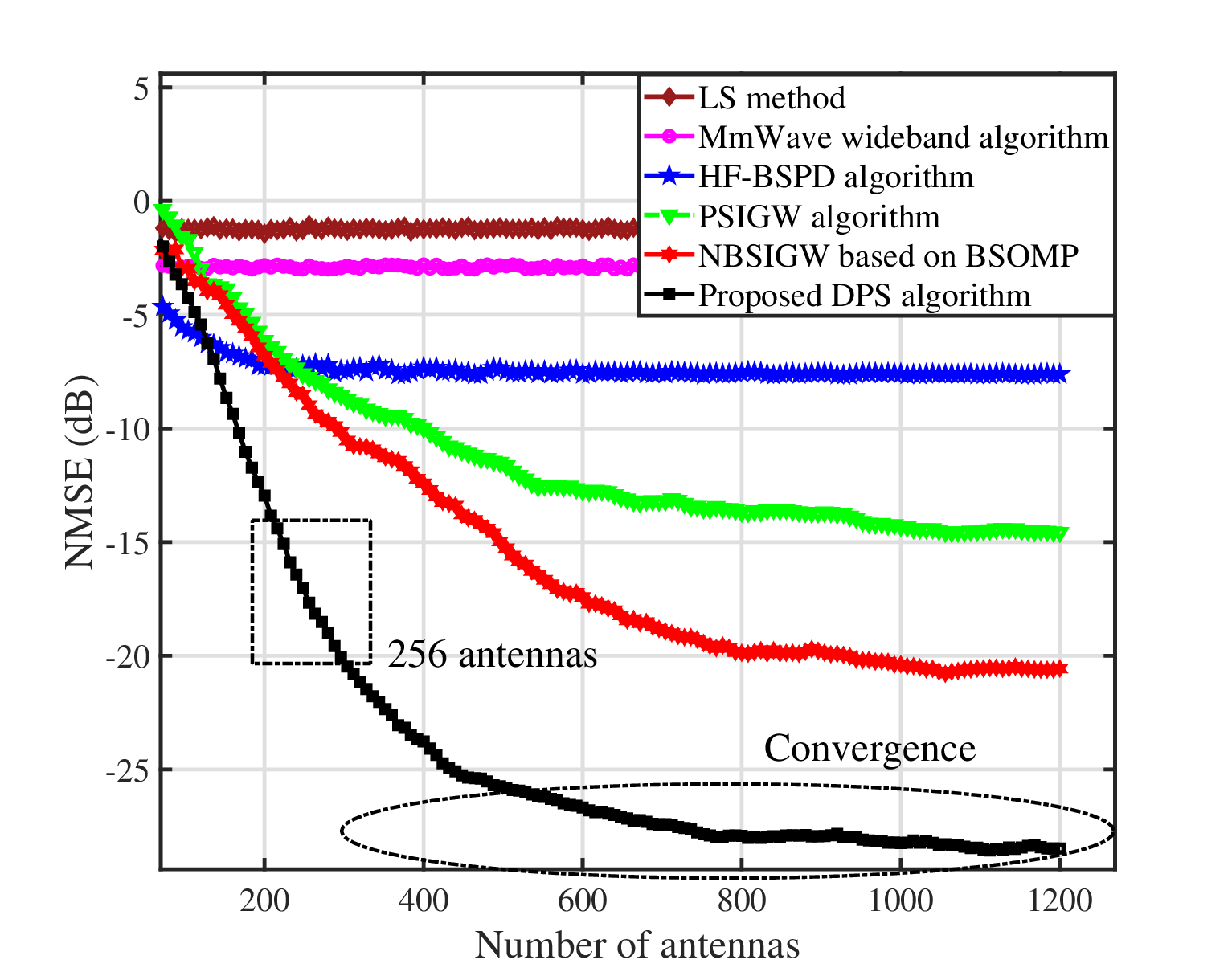}
  \caption{NMSE against the number of antennas, where $N_{\rm s} = 4$.} \label{Fig: NMSE against N}
\end{figure}

\begin{figure}
  \centering
  \includegraphics[scale=0.32]{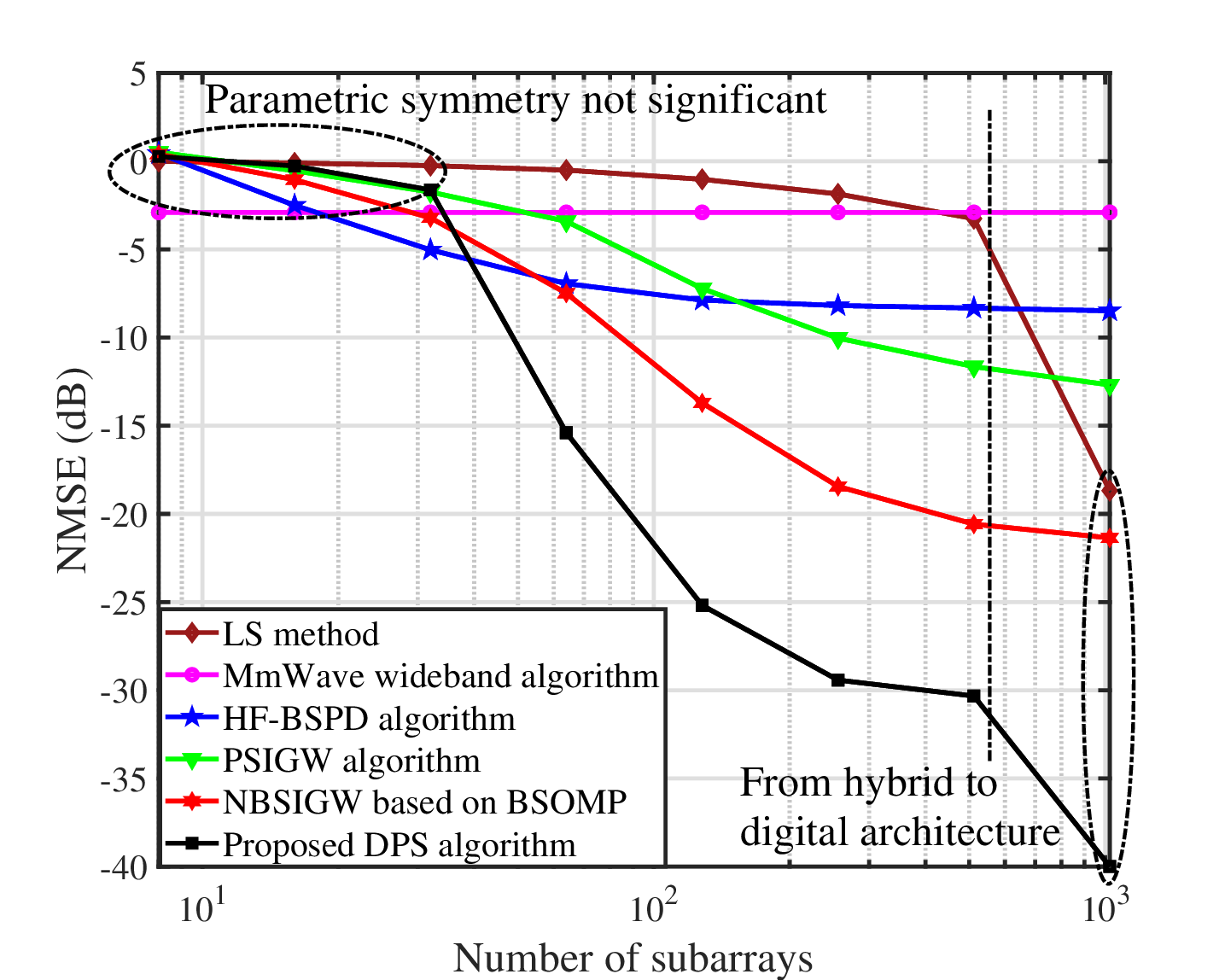}
  \caption{NMSE against the number of subarrays, where $N$ = 1024.} \label{Fig: NMSE against K}
\end{figure}

Then, the NMSE against $N$ is evaluated, where the SNR is 10 dB, and $N_{\rm s} = 4$. As shown in Fig.~\ref{Fig: NMSE against N}, the NMSE decreases as $N$ increases, since a higher-dimensional signal can be received. Although the PSIGW and NBSIGW algorithms can both achieve accurate estimation results, they have high computational complexity when $N$ is large. The proposed DPS algorithm achieves the lowest NMSE compared to that of other algorithms, and maintains a low computational complexity even when $N$ is large. Meanwhile, as $N$ increases to 512, the NMSE tends to converge to -30 dB. The NMSE against the number of subarrays is also evaluated as in Fig.~\ref{Fig: NMSE against K}, where ${\rm SNR} = 10$ dB, and $N = 1024$. The proposed DPS algorithm has a lower NMSE than that of other algorithms, while the NMSE improves rapidly as $K$ increases. Note that the best NMSE can be obtained when $K = N$, in which case the XL-MIMO system is equipped with a fully digital precoding architecture. However, the NMSE will be compromised if only a limited number of subarrays are available. A limited number of delays observed by the CPU will result in the estimate being easily affected by the noise. Meanwhile, the assumption in \eqref{Eq: Section2_16} becomes invalid due to the larger apertures of the subarrays. In cases where $K$ is small, the conventional CS-based algorithms can be used alternatively to estimate the channel.

\begin{figure}
  \centering
  \includegraphics[scale=0.33]{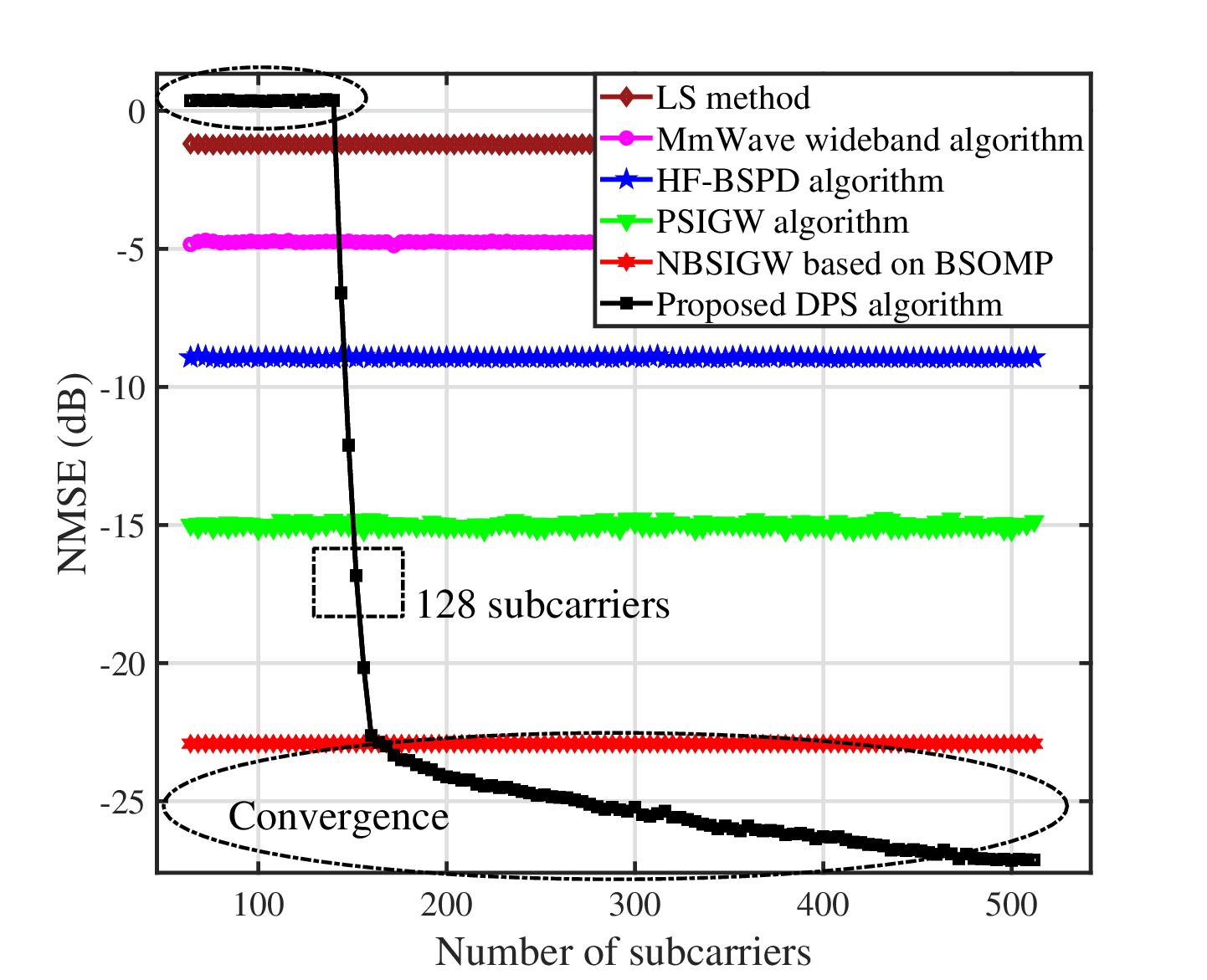}
  \caption{NMSE against the number of subcarriers, where $B = 600$ MHz.} \label{Fig: NMSE against M}
\end{figure}

Moreover, the NMSE against the number of subcarriers is evaluated, where $B = 600$ MHz, and $K = 256$. As shown in Fig.~\ref{Fig: NMSE against M}, the NMSE of the reconstructed channel will be compromised when the value of $M$ is small. In such a scenario, the delay variation across the entire array is smaller than the delay resolution. All the subarrays have the same delay estimate, and thus the parametric symmetry becomes insignificant. Consequently, in the event that all LPUs estimate the same delay from ${\bf F}_{\tau}$, the CPU will adaptively switch to conventional CS-based algorithms for channel estimation. It is important to note that the NMSE exhibits a rapid decrease as $M$ increases, and the NMSE tends to converge when $M$ exceeds 256. It is evident that the proposed DPS algorithm does not require a significant number of subcarriers for precise channel estimation, which can facilitate a further reduction in complexity and conserve the frequency resources.

\begin{figure}
  \centering
  \includegraphics[scale=0.36]{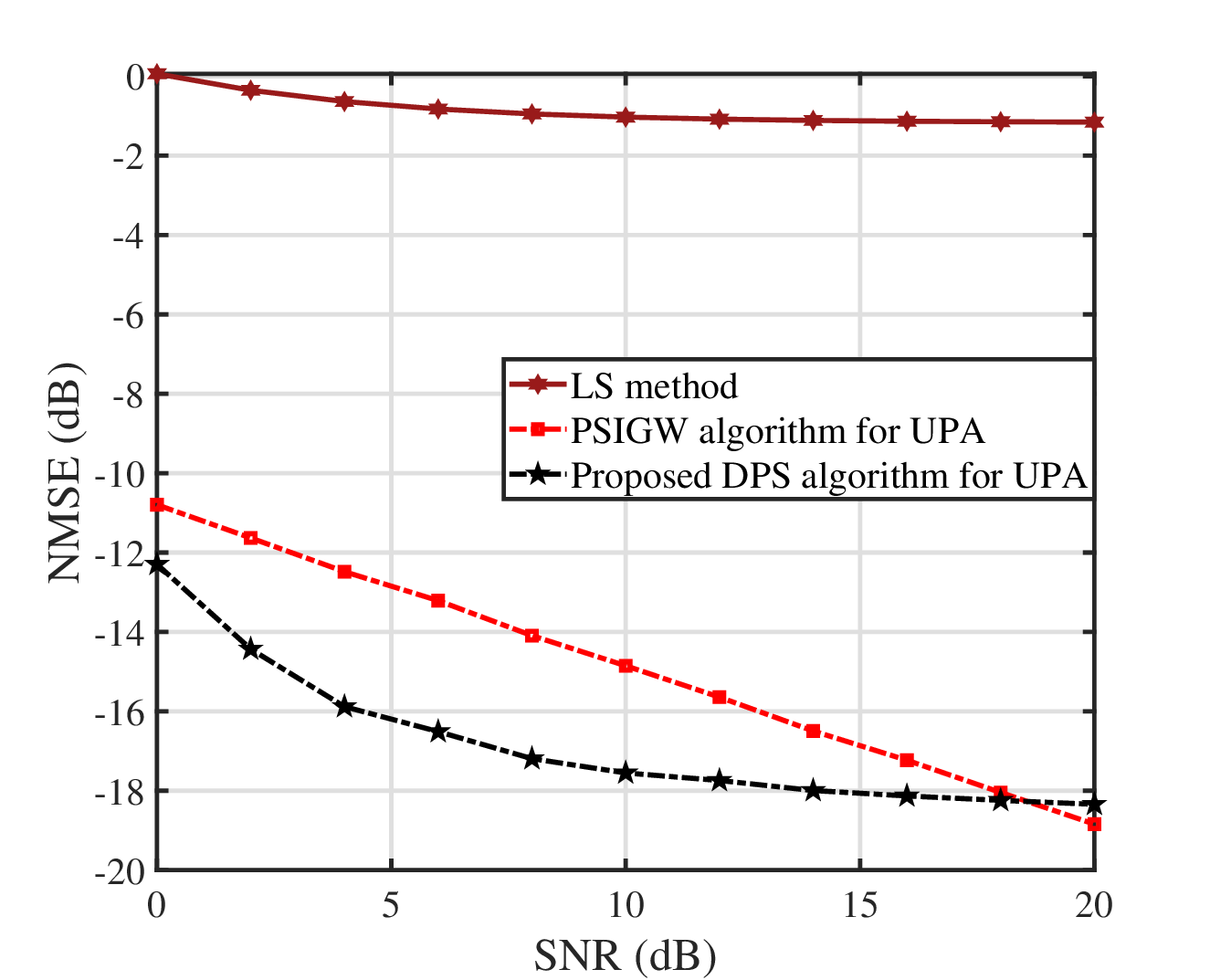}
  \caption{NMSE of the reconstructed wideband near-field channel (UPA).} \label{Fig: NMSE of UPA}
\end{figure}

Finally, the NMSE of the proposed DPS algorithm for XL-MIMO systems with UPAs is evaluated, where $N_{\rm r} = 8$, $N_{\rm c} = 256$, $M = 256$, and $K = 256$. As shown in Fig.~\ref{Fig: NMSE of UPA}, the wideband near-field channel can still be accurately estimated when the BS is equipped with a UPA, and the NMSE is lower than that of other algorithms. However, the NMSE is higher than that in XL-MIMO systems with ULAs, since the parametric symmetry diminishes as the array aperture reduces.

\section{Conclusion}\label{Sec: Conclusion}

In this paper, a U6G hybrid XL-MIMO system with a distributed processing architecture was considered. In order to accurately estimate the channel under the beam squint effect and with a distributed hybrid architecture, the parametric symmetry of the wideband near-field channels was explored. Based on this, a DPS algorithm involving only a single pilot, was proposed. The delay of the central subarray was estimated in the central LPU, while the delays of other subarrays were extrapolated across the LPUs. Then, based on the parametric symmetry, the channel parameters were decoupled and individually estimated at the CPU, by only linearly combining the estimates from LPUs with low computational complexity. Finally, the complex gains were calculated in the LPUs to reconstruct the channel. Moreover, the computational complexity, the multiple-path resolution, the CRLBs and LBs of the estimates in hybrid XL-MIMO systems and in distributed processing architectures, were evaluated, respectively, which indicated that the proposed DPS algorithm performs well in a scenario with extremely large arrays and bandwidths. Our simulation results proved that the proposed DPS algorithm can reconstruct the channel accurately, while requiring only a single pilot and a much lower complexity.

\begin{appendices}
\section{}

\begin{figure*}[ht]
\centering
\begin{equation}\label{Eq: SectionA_1}
{\bf U}_{\rm dh} = \left[ {\begin{array}{*{20}{c}}
{\frac{{K\left( {{K^2} - 1} \right){s'^2}\left( {20{d^2} + \left( {3{K^2} - 7} \right){s'^2}{\theta ^2}} \right)}}{{240{d^2}}}}&{\frac{{K\left( {{K^2} - 1} \right){s'^2}\theta \left( {\left( {3{K^2} - 7} \right)\left( {1 - {\theta ^2}} \right){s'^2} - 40{d^2}} \right)}}{{480{d^3}}}}&{ - \frac{{K\left( {{K^2} - 1} \right){s'^2}\theta }}{{12d}}}\\
{\frac{{K\left( {{K^2} - 1} \right){s'^2}\theta \left( {\left( {3{K^2} - 7} \right)\left( {1 - {\theta ^2}} \right){s'^2} - 40{d^2}} \right)}}{{480{d^3}}}}&\frac{{K\left( {{K^2} - 1} \right)\left( {1 - {\theta ^2}} \right){s'^2}\left( {\left( {3{K^2} - 7} \right)\left( {1 - {\theta ^2}} \right){s'^2} - 80{d^2}} \right)}}{{960{d^4}}}&{K - \frac{{K\left( {{K^2} - 1} \right){s'^2}\left( {1 - {\theta ^2}} \right)}}{{24{d^2}}}}\\
{ - \frac{{K\left( {{K^2} - 1} \right){s'^2}\theta }}{{12d}}} & {K - \frac{{K\left( {{K^2} - 1} \right){s'^2}\left( {1 - {\theta ^2}} \right)}}{{24{d^2}}}} & K
\end{array}} \right]
\end{equation}
\hrule
\end{figure*}

Substituting \eqref{Eq: Section4_9} into \eqref{Eq: Section4_13}, the coefficient matrix ${\bf U}_{\rm dh}$ can then be further expressed as \eqref{Eq: SectionA_1}, which is derived by $\sum \limits_{k = 1}^{K}1 = K$, $\sum \limits_{k = 1}^{K}\delta_{K,k} = 0$, $\sum \limits_{k = 1}^{K}\delta_{K,k}^2 = \frac{K\left(K^2-1\right)}{12}$, $\sum \limits_{k = 1}^{K}\delta_{K,k}^3 = 0$, and $\sum\limits_{k=1}^K\delta_{K,k}^4 = \frac{K\left(K^2-1\right)\left(3K^2-7\right)}{240}$, respectively, where $s' = N_{\rm s}s$. Then, the determinant of the coefficient matrix ${\bf U}_{\rm dh}$ can be calculated according to
\begin{equation}\label{Eq: SectionA_2}
\begin{aligned}
&{\rm det}\left({\bf U}_{\rm dh}\right) = \left|{\bf U}_{\rm dh}\right| = {N_{\rm s}^6 s^6}{K^3} \times \\ &\frac{{{\left( {{K^2} - 1} \right)}^2}\left( {1 - {\theta ^2}} \right)\left( {2{K^2} - 8 + 7{K^2}{\theta ^2} - 13{\theta ^2}} \right)}{{17280{d^4}}} \\ \approx &\frac{N_{\rm s}^6 s^6 K^9 \left(1-\theta^2\right) \left(2 + 7\theta ^2\right)}{17280 d^4},
\end{aligned}
\end{equation}
which is given by $K^2-1 \approx K^2$, and $2K^2-8 + 7K^2\theta^2-13\theta^2 \approx 2K^2-7K^2\theta^2$, since $K \gg 1$ in XL-MIMO systems; ${\rm det}\left(\cdot \right)$ denotes the determinant operation. Then, the diagonal elements of ${\bf U}_{\rm dh}^* $ (which represent the adjoint matrix of ${\bf U}_{\rm dh}$) can be calculated as
\begin{equation}\label{Eq: SectionA_3}
\begin{aligned}
{{\left[ {{\bf{U}}_{\rm{dh}}^ * } \right]}_{1,1}} = &\frac{{K^2}\left( {{K^2} - 1} \right)\left( {{K^2} - 4} \right){N_{\rm s}^4s^4}{{\left( {1 - {\theta ^2}} \right)}^2}}{720{d^4}} \\ \approx &\frac{{K^6}{N_{\rm s}^4s^4}{{\left( {1 - {\theta ^2}} \right)}^2}}{720{d^4}},
\end{aligned}
\end{equation}
and
\begin{equation}\label{Eq: SectionA_4}
\begin{aligned}
{{\left[ {{\bf{U}}_{\rm{dh}}^ * } \right]}_{2,2}} = &\frac{{K^2}\left( {{K^2} - 1} \right){N_{\rm s}^2s^2}\left( {15{d^2} + {N_{\rm s}^2s^2}{\theta ^2}\left( {{K^2} - 4} \right)} \right)}{180{d^2}} \\ \approx &\frac{{K^4}{N_{\rm s}^2 s^2}\left( {15{d^2} + {N_{\rm s}^2s^2}{\theta ^2} {K^2}} \right)}{180{d^2}},
\end{aligned}
\end{equation}
and
\begin{equation}\label{Eq: SectionA_5}
\begin{aligned}
&{\left[ {{\bf{U}}_{\rm{dh}}^ * } \right]_{3,3}} \approx {K^4}N_{\rm s}^2{s^2} \times \\ &\frac{960{d^4} + {K^2}{N_{\rm s}^2s^2}{d^2}\left( {144{\theta ^2} - 80} \right) + 3{K^4}{N_{\rm s}^4s^4}{{\left( {1 - {\theta ^2}} \right)}^2}} {{11520{d^4}}},
\end{aligned}
\end{equation}
which stem from $K^2-4 \approx K^2$, and $3K^3-7 \approx 3K^2$, since $K \gg 1$ in XL-MIMO systems. Hence, the CRLBs of $\left\{ \theta,d,r\right\}$ can be denoted as
\begin{equation}\label{Eq: SectionA_6}
\begin{aligned}
&\left\{ {\theta_{\rm CB},d_{\rm CB},r_{\rm CB}} \right\} = \left(\frac{c\sigma}{\pi \left| g \right|\sqrt{P}}\right)^2 \times \\ &\frac{3}{2N_{\rm s}M^3\Delta f^2} \left\{ {\frac{{\left[ {{\bf U}_{\rm fd}^ * } \right]}_{1,1}}{\left| {{\bf U}_{\rm fd}} \right|},\frac{{\left[ {{\bf U}_{\rm fd}^ * } \right]}_{2,2}}{\left| {{\bf U}_{\rm fd}} \right|},\frac{{\left[ {{\bf U}_{\rm fd}^ * } \right]}_{3,3}}{\left| {{\bf U}_{\rm fd}} \right|}} \right\},
\end{aligned}
\end{equation}
where $M^2 -1 \approx M^2$, since $M \gg 1$ in wideband systems. Then, the CRLBs of $\left\{\theta,d,r\right\}$ can be expressed as \eqref{Eq: Section4_16}, where $N = KN_{\rm s}$. The proof is concluded. \qed

\end{appendices}

\end{document}